\documentclass[a4paper,fleqn,usenatbib,referee,epsfig,color]{mnras}

\usepackage[T1]{fontenc}

\usepackage{ae,aecompl}

\usepackage{graphicx}	% Including figure files
\usepackage{amsmath}	% Advanced maths commands
\usepackage{amssymb}	% Extra maths symbols

\newcommand{\be}{\begin{equation}}
\newcommand{\ee}{\end{equation}}
\newcommand{\bea}{\begin{eqnarray}}
\newcommand{\eea}{\end{eqnarray}}

\newcommand{\A}{\bar A}

\title[ halo magnetic spiral arms]{\bf Magnetic Spiral Arms in Galaxy Halos}

\author[R.N. Henriksen]{R.N. Henriksen$^1$\thanks{henriksn@astro.queensu.ca}\\
$^1$Dept. of Physics, Engineering Physics \& Astronomy, Queen's University, Kingston, Ontario, K7L 3N6, Canada}

\date{Accepted XXX.  Received YYY; in original form ZZZ }

\pubyear{2016}

\begin{document}
\label{firstpage}
\pagerange{\pageref{firstpage}--\pageref{lastpage}}
\maketitle

\begin{abstract}
We seek the conditions for a {\it steady}  mean field galactic dynamo. The  parameter set  is reduced to those appearing in the $\alpha^2$ and $\alpha/\omega$ dynamo, namely velocity amplitudes, and  the ratio of sub-scale helicity to diffusivity.  The parameters can be allowed to vary on conical spirals. We analyze the mean field dynamo equations in terms of  scale invariant logarithmic spiral modes and special exact solutions.  Compatible scale invariant gravitational spiral arms are introduced  and illustrated in an appendix,  but the detailed dynamical interaction with the magnetic field is left for another work. As a result of planar magnetic spirals `lifting' into the halo,  multiple sign changes in average rotation measures  forming a regular pattern on each side of the galactic minor axis,  are predicted. Such changes have recently been detected in the CHANG-ES survey. 
\end{abstract}


\section{Introduction}

Magnetohydrodynamic dynamo theory has  a long history and a rather rigorous foundation. Reviews of early pioneering fundamental work may be found in \citep{RS1992,K1999} and  for astrophysical applications in \citet{RSS1988}. 

The theory has found perhaps its most testable astrophysical application in the explanation of large scale magnetic fields in spiral galaxies (e.g. \citet{CSSS2014}). The extended, quasi-transparent nature of these objects allows the components of the galactic dynamo to be observed. This is especially true when a combination of face-on and edge-on spiral galaxies (e.g. \citet{Beck2016} and \citet{Kr2015}) is considered. 
The systematic study of a sample of edge-on galaxies in the CHANG-ES survey \citet{WI2015}  has reinforced the belief in an  intimate connection between halo and disc magnetic fields. Such a connection has long been advocated for external galaxies by Krause (e.g.\citet{Kr2015}), and it has been brilliantly demonstrated recently for the Milky Way (\citet{Gfarr2015}). 

That lagging haloes may be an indication of a magnetic connection to the galactic environment, has recently been argued in \citet{HI2016} for a special case ($v$ equal to the Alfv\'en speed) of the axisymmetric  mode. There is in that work a kind of simple `dynamo' acting, which is effected by shearing the magnetic field between the disc and the  distant environment.  Such lags may also be produced by classical  dynamo action.  We find  here that  dynamo action can (but not always)  produce magnetic spirals on conical surfaces that extend into the halos of galaxies with decreasing amplitude, just as in \cite{HI2016}. If the velocity field is simply proportional to the magnetic field, then lags will arise.  The dynamo action will also be revealed by oscillations in the sign of rotation measures and/or in the oscillation of polarization intensities, particularly when this occurs in a regular pattern on the same side of the galactic minor axis. 
 
The observations of face-on galaxies imply the existence of magnetic spiral arms (e.g. \citet{Kr1993}, \citet{BH1996} and \citet{Beck2015}). There has been an intensive effort to explain the origin of these arms (e.g. \citet{Moss2013}, \citet{CSS2015} and references therein) including their location, which is often different from that of the gravitational spiral arms. This effort has been conducted largely numerically, but it has led to a number of innovative physical ideas. 

These physical ideas involve among others  the variation of vertical outflow and turbulence with respect to the gravitational spiral arms, the notion of a relaxation delay in the `alpha' effect, and the disc/halo interaction. The numerical models have also had to deal with  `quenching' of the turbulent alpha effect and dynamical effects due to gravity and magnetohydrodynamical back reaction. Despite these efforts, there does not seem to be a consensus as yet as to the physical origin of the magnetic spiral arms, in part due to the rich collection of parameters.

The present work does not attempt to deal directly with these questions. Rather it provides a simple model for the existing  magnetic spiral arms in the hope of minimizing the number of  parameters essential to their existence. We assume a steady state and we look for spiral arm  solutions by expanding the magnetic field in logarithmic spiral modes. This avoids the `quenching' problem associated with unlimited growth of the field, and by insisting on spiral modes, addresses directly the conditions necessary for their existence.   An expansion of these fields into galactic halos is found, although the description is limited to moderate heights above the disc in our general approach.  An exact analytic solution extends this behaviour to arbitrary heights. 

%The development of the steady state from initial conditions concerns the evolution of the galactic disc, and indeed of the galaxy itself. We leave this difficult phase to a related work, in which the time dependence will be subjected to the same mathematical approach. 

%Our approach is not perturbative, but rather finds scale invariant solutions . This  allows us to infer the spatial form of the alpha coefficient, of the turbulent diffusion coefficient, and of the three dimensional velocity, in order that  scale-free spiral magnetic arms exist. The solutions are however limited to regions in which the ratio $z/r$ is small. 

There are some physical gaps  due to our assumptions. Because of the assumed steady state and the modal analysis, there is always an arbitrary constant in our solutions. Hence we can not predict the strength of observed fields, although in principle one can try to fit them globally (or locally in a peculiar region) with that one parameter. We clearly can not discuss growth rates of the magnetic field. Moreover we are forced to discuss the axisymmetric mode (both magnetically and gravitationally) separately, because of the form of the scale invariance assumed. However such behaviour can be added when necessary, and is studied for example in \citet{HI2016}.

In addition we have  to ask in what reference frame is the magnetic spiral steady. We will usually assume that it is steady in a frame rotating rigidly with a pattern angular velocity $\Omega$, which might be that of the gravitational arms should these have a  distinct pattern speed. Subsequently, the rotation velocity would be added to the dynamo velocity to yield the observed velocity. In the local inertial or `systemic' frame, the boundary condition on the dynamo velocity is that of the rotating disc at the disc and zero at infinity.  In the disc frame the  boundary conditions would be rather the reverse, zero at the disc and negative disc velocity at infinity. When the velocity ${\bf v}$ is taken parallel to the magnetic field $ {\bf B}$, we suppose that both quantities are in the pattern frame, which may also be  the systemic frame.   

Unless  the pattern rotation is zero, the magnetohydrodynamics should also be analyzed in the rotating frame. The simplest possibility is to take the peculiar mean velocity in the rotating frame to be Alfv\'enic and the halo gas to be incompressible \citet{HI2016}). A macroscopic equilibrium due to pressure, magnetic, gravitational and inertial forces can then  be found straightforwardly. The incompressibility assumption is quite reasonable if the cosmic ray (CR) pressure is dominant in the halo, since the sound speed ($c/\sqrt{3}$) implies a wave crossing time that is short on a galactic time scale. This is still true for the hot halo component. However we will not discuss the rotational dynamics in detail in this work. Some discussion of Faraday's law in a rotating frame is required however, since this is at the origin of the dynamo equation. 

Ignoring explicit gas dynamics is a means of surveying rapidly the effects of different flow topologies on dynamos. Although scale invariance is often the asymptotic state of a complex physical system, no rigorous prediction for the velocity field should be expected in this survey. For simplicity we have used a scale invariant class ($a=2$) that corresponds to conservation of specific angular momentum. This implies the somewhat uncomfortable $1/r$ velocity and magnetic field dependence in a pattern frame.It would be more realistic in future work to use the class $a=1$, which implies a velocity and magnetic field constant in radius in the pattern frame of reference. If however the vertical scale height is much smaller than the radial disc scale, then even for $a=2$ the velocity tends to a constant (e.g. appendix of \citet{Hen2017},$\alpha=\delta_\perp$).

The previous work most similar to our own in spirit is that of \citet{SM1993}, since they assume a steady state in the pattern reference frame.  However it is reassuring that previous numerical work on the time dependent version of our equations has uncovered similar results. These are found for example in \cite{DB90} and \cite{MBDT93}, which are discussed briefly with our conclusions. The steady state and scale invariance together replace many of the parameters adopted in these papers. 

%We differ from that work essentially by using scale-invariance to find  modal solutions, which include an approximate but consistent $z$ dependence. Moreover we emphasize the advantage of expanding in terms of spiral magnetic modes `ab initio'. Our model is steady so that we infer only a spatial dependence of the velocity, which is dictated by the scale invariance.

 In the next section we formulate the equations of the problem under our assumptions. This includes the modal and scale invariant ans\"atze. In the following section we find the conditions on physical quantities  required for an exact scale invariant example. Subsequently we illustrate the steady magnetic dynamo field in and above the galactic disc for several  cases depending on the nature of the velocity.    

In appendix A we discuss at length the exact scale-invariant equations expressed both in terms of the vector potential and of the magnetic field. These are used to show the transition to the approximate equations that we use for modal analysis in the text. 

 In Appendix B we introduce the  {\it scale invariant} Kalnajs \citep{Kal1971}  model  of the gravitational spiral arms, and indicate how these might eventually be coupled to the magnetic spiral arms.  Illustrative figures are included in that appendix, as is also a speculation regarding the separation of gravitational and magnetic arms.  
     
\section{Kinematic Dynamo Theory}

It is worth remembering that the classical isotropic mean-field dynamo theory relies on an inertial 
electric field ${\bf E}$ that is due to flux-freezing, plus subscale `turbulent' diffusion and generation. This takes the form in Gaussian Units (e.g. \citet{M1978})
\be
{\bf E}=-\frac{\bf v}{c}\wedge {\bf B}+\frac{(\eta-\beta_d)}{c}\nabla\wedge {\bf B}-\frac{\alpha_d}{c}{\bf B},\label{eq:Efield}
\ee  
where ${\bf B}$ is proportional to the mean magnetic field and ${\bf v}$ is the mean velocity field. It is convenient to regard the vectors ${\bf E}$ and ${\bf B}$ to be the true mean fields divided by $\sqrt{4\pi\rho}$, where $\rho$ is some fiducial gas density. Accordingly, each of these vectors has the Dimension of velocity in electromagnetic Units (emu). We have labelled the subscale alpha and beta effects by $\alpha_d$ and $\beta_d$ respectively, but subsequently we absorb the beta effect into the subscale diffusion by letting 
\be
\eta-\beta_d\leftarrow \eta.\label{eq:neweta}
\ee 
Note that there is no electrostatic field included, since that would require charge separation.

In a true steady state in an inertial frame of reference the integral of the electric field around an arbitrary stationary contour is zero by Faraday's law. Consequently the electric field itself may be taken zero, but for an unlikely electrostatic term. After introducing the vector potential ${\bf A}$ and multiplying equation (\ref{eq:Efield}) by $-c$, we obtain our working equation in the form
\be
0={\bf v}\wedge \nabla\wedge {\bf A}-\eta\nabla\wedge\nabla\wedge {\bf A}+\alpha_d\nabla\wedge {\bf A}.\label{eq:Afield}
\ee
This is equivalent to the standard form  involving the curl of this equation, provided that the electrostatic field is neglected. 

One should note that setting equation (\ref{eq:Efield}) equal to zero yields the dynamo equations directly in terms of the magnetic field. These equations do not automatically contain the solenoidal constraint unless the substitution in terms of vector potential is made.

The above argument holds strictly only in an inertial frame. In a frame rotating with the pattern speed (primed), Faraday's law becomes, to first order in $v/c=\Omega r/c$, 
\be
\nabla\wedge {\bf E}'\approx -\frac{1}{c}\partial_t{\bf B}\mid_{(r,\phi',z)}+\frac{\Omega}{c}(B_r\hat{\bf e}_\phi-B_\phi\hat{\bf e}_r),\label{eq:newfaraday}
\ee
where we have moved $-\nabla\wedge ({\bf v}\wedge {\bf B})$ to the right hand side of the equation and calculated it explicitly there.
Thus, so long as $c/\Omega$ is much larger than a typical scale in the galactic disc and halo, we can neglect the new term on the right compared to the curl of the electric field.  The argument above will thus apply approximately to a steady state in a uniformly rotating frame. One assumes  that equation (\ref{eq:Efield}) and hence equation (\ref{eq:Afield}) holds  in the rotating frame.
 
We proceed by seeking a scale invariant solution for the vector potential. Following the procedure advocated in \citet{Hen2015}. We use the symbols $\alpha$ and $\delta$ for reciprocal temporal and spatial scales, each of which may be taken to have Dimension of reciprocal length in a steady state. Then we transform from cylindrical coordinates $\{r,\phi,z\}$ (referred to the galactic axis) to self-similar or 'Lie invariant' coordinates $\{R,\kappa,Z\}$ according to 
\be
\delta r\equiv e^{\delta R},~~~~\delta z\equiv Ze^{\delta R},~~~~\kappa\equiv q\phi+\delta R+iZ\equiv \xi+i\frac{z}{r}.\label{eq:trans1}
\ee
We note that 
\be
Z\equiv \frac{z}{r},\label{eq:Z}
\ee
so that it is constant on cones whose generators pass through the centre of the galaxy.

The quantity $\xi$ may also be written as 
\be
\xi\equiv q\phi+\ln{\delta r},\label{eq:logspiral}
\ee
which shows each constant value of $\xi$ to define a logarithmic spiral having $q$ as the tangent of its (negative)  pitch angle. If $q$ is positive the spirals are trailing relative to the positive sense of $\phi$. This sense is that of the pattern speed, normally given by the rotation of the galaxy.  One can describe trailing spirals  with $q<0$ by taking the sense of increasing $\phi$  to be opposite to the sense of galactic rotation.

Strictly speaking our transformed coordinates should be $\{R,\xi,Z\}$, but the scale invariance reduces them to partial differential equations in $\xi$ and $Z$. We give these equations in their exact form  in appendix  A for possible future use.  We also give the equations in terms of the magnetic field in the Appendix, because these allow exact `toy' models to be found.

% Because equation (\ref{eq:Afield}) contains only the curl of the vector potential, the equations can also be expressed in terms of the mean magnetic field.

 Our approximate treatment lies in merging the separate  $Z$ and $\xi$ dependence  into $\kappa$  as in equation (\ref{eq:trans1}) and ignoring any other $Z$ dependence. This is reasonably accurate so long as $z/r$ is small.  We  have inserted the square root of  $-1$ (i.e. $i$) in the $Z$ dependence of the scale invariant quantity $\kappa$, in order that our spiral modes (see subsequently) may have an exponential decline (or increase) with $z$. If $i$ is suppressed each node will have a  periodic behaviour in $z$ and it would require a Fourier sum to obtain a reasonable behaviour in $z$. Although this remains formally a viable option, the exponential decline  (and occasionally growth) with distance from the disc for each node at fixed $r$ seems more physical and indeed appears naturally in sample analytic solutions.  We note that the  vertical scale height of the magnetic field is given by the radius at each point.

According to the scale invariant ans\"atz, we are required to assign a scaling to each physical variable depending on its Dimension. For example, the Dimension co-vector of the vector potential is ${\bf d}_A=(-1,2)$ in the $\{\alpha,\delta\}$ scaling space. This gives its exponential scaling factor as $(-\alpha+2\delta)$ so that \citep{Hen2015}
\be
~~~~~~~~~~~~~~~~~~~~~~~~~~~~~~{\bf A}=\bar{\bf A}(R,\xi,Z)e^{(2\delta-\alpha)R}.\label{eq:Aform}
\ee 
For scale invariance, the dependence on $R$ in the scaled vector potential $\bar{\bf A}$ must not appear, since we have chosen it to be along the (Lie) symmetry direction \citep{Hen2015}. Taking into account the various Dimensional co-vectors \footnote{The Dimension of $\alpha_d$ is the same as that of a velocity while that of $\eta$ is the same as that of specific angular momentum and of {\bf A}} and summarizing, our scale invariant ans\"atz is 
\bea
{\bf A}&=&\bar{\bf A}(\kappa)e^{(2-a)\delta R},\label{eq:Aformsum}\\
{\bf v}&=&\bar{\bf v}(\kappa)e^{(1-a)\delta R},\label{eq:vformsum}\\
\alpha_d&=&\bar\alpha_d(\kappa)e^{(1-a)\delta R},\label{eq:alphaeff}\\
\eta&=& \bar\eta(\kappa)e^{(2-a)\delta R}.\label{eq:diffeff}
\eea
Here we have set the ratio of the arbitrary scales equal to $a$, that is 
\be
~~~~~~~~~~~~~~~~~~~~~~~~~~~~~~~~~a\equiv \frac{\alpha}{\delta},\label{eq:Sclass}
\ee
which defines the self-similar `class'\citep{CH1991,Hen2015}. 

%It does not seem to be possible  for $m\ne 0$ to introduce explicit anisotropic spatial scaling to the scale invariant Symmetry(involving say $Z$ with a different scaling) as in the Blasius problem discussed in \citet{Hen2015}). An anisotropy does arise naturally under the scale invariant Symmetry however, since the $z$ dependence is weighted by the radius (see equation \ref{eq:Aexp} below). 

%We might impose anisotropic spatial scales $\delta=(\delta_\perp +\delta_\parallel)/2$, where $\delta_\perp$ and $\delta_\parallel$ refer to vertical and parallel scaling respectively. This changes nothing in our subsequent discussion unless we also suppose 
%\be
%\alpha_d=\bar\alpha_d e^{(\delta_\perp-\alpha)R}\label{eq:alphagen}
%\ee
%in place of equation (\ref{eq:alphaeff}).  Such a dependence on the vertical scaling is suggested by the notion of rising helical turbulence. The advantage of this kind of anisotropy is that it permits a more general radial dependence of $\alpha_d$ on $r$ according to (maintaining equation (\ref{eq:trans1}), especially the radial variable)
%\be 
%\alpha_d=\bar\alpha_d (\delta r)^{\frac{\delta_\perp-\alpha}{\delta}},
%\ee

%which may be compared to equation (\ref{eq:alphaeff}) above or equivalently (\ref{eq:alphaexp}) below. The radial dependence of other quantities could be generalized in a similar fashion if one were to define the similarity class as in terms of say $\alpha/\delta_\perp$ in the scale invariant ans\"atz, but we mostly  avoid this complication here.

We proceed here with  spatially isotropic scaling, but different scalings parallel and perpendicular to the disc are possible (e.g. Appendix in \citet{Hen2017}). It allows slightly generalized power laws to be the radial power law dependences.

There remains now  some tedious algebra in order to transform equation (\ref{eq:Afield}) into a useful form using the scale invariant quantities. We give the result  for general $a$  (after approximation but see appendix A for the complete equations) for reference, but we do not analyze the general case in this work. The equations have only one physical parameter in addition to the similarity class $a$ and the velocity components ${\bf v}=\{u,v,w\}$. We take this parameter to be 
\be
\Delta\equiv \frac{\bar\alpha_d}{\delta\bar\eta},\label{eq:parameter}
\ee 
which is essentially a Reynolds number ( of the sub-scale turbulence) and is essentially the dynamo number used elsewhere \cite{B2014}. The velocity components $\{u,v,w\}$ are now in Units of $\bar\eta\delta$, that is 
\be
\bar{\bf v}=(\bar\eta\delta)(u,v,w).\label{eq:scaledvel}
\ee 

 Letting a prime denote differentiation with respect to $\kappa$ the   steady dynamo equations for the vector potential components $\{\bar A_R,\bar A_\phi,\bar A_Z\}$ become  at sufficiently small $Z$ (see  Appendix A)
\bea
0&=&(q^2-1)\A''_r-(qv+iw~ s1)\A'_r-q\A''_\phi+(v-(3-a)q-i\Delta ~s1)\A'_\phi+(3-a)v\A_\phi\nonumber\\
&-& i\A''_z~ s1+(w+q\Delta+(a-1)s1)\A'_z+(2-a)w\A_z,\label{eq:dynamor}\\
0&=& -q\A''_r+(qu-q(1-a)+i\Delta s1)\A'_r-(u-2(2-a)+iw~ s1)\A'_\phi+(3-a)(1-a-u)\A_\phi\nonumber\\
&-&iq\A''_z~ s1-(\Delta-qw)\A'_z-\A_z(2-a)\Delta,\label{eq:dynamophi}\\
0&=& -i\A''_r~ s1-s1~(q\Delta~ s1+i(2-a)-iu)\A'_r-iq\A''_\phi ~s1+(\Delta+iv~ s1)\A'_\phi+\A_\phi(3-a)\Delta\nonumber\\
&+&(1+q^2)\A''_z-(u+qv-2(2-a))\A'_z-(2-a)(u-(2-a))\A_z,\label{eq:dynamoz}
\eea
where
\be
~~~~~~~~~~~~~~~~~~~~~~~~~~~~~~~~~~~ s1\equiv sgn(Z)sgn(m)sgn(q). \label{eq:expswitch}
 \ee

We  proceed by taking the dependence on $\kappa$ to be  the modal form 
\be
{\bf \A}={\bf C}e^{ip\kappa}\equiv {\bf C}\exp{(ip\xi-pZs1))},\label{eq:C}
\ee
where ${\bf C}$ is a constant vector and $p\equiv m/q$. We note that if we wish the  magnetic field to decrease with increasing $z$, then $s1$ and $m$ should have the same sign above the plane and opposite signs below the plane. The reverse sign assignment will produce a field increasing into the halo, which we do not exclude ab initio.

%If equations (\ref{eq:Afield}) are written out in full in terms of $\xi$ and $
%Z$, then terms explicitly proportional to $Z$ appear. The order of magnitude of the derivatives may be estimated from $\partial_Z=O(p)$ and $\partial_\xi=O(ip)$.  
In order to reduce the problem to this set of three ordinary equations for the vector potential as a function of $\kappa$, approximation has been necessary. All terms proportional to $Z$ occurring in the  equations (\ref{eq:Afield})  if written out in full in terms of the variables  $\xi$ and $Z$,  are neglected (see Appendix A).  Moreover a term in $\partial_Z\A_z$ has been neglected in the diffusion part of the radial equation (see AppendixA).

One can allow  the diffusion coefficient  and  the Reynolds parameter to depend on $\kappa$, provided the functional dependence is the same for each. That would allow azimuthal anisotropy in the turbulence e.g \citet{SM1993}. If $\bar\alpha_d$ and $\bar\eta$ have the same dependence on $\kappa$, the equations will not be affected  since $\Delta$ remains  constant.  If the velocity terms are present, equation (\ref{eq:scaledvel}) shows that in order for the velocity parameters $\{u,v,w\}$ to be constant (which leaves the equations unchanged), one must have the scaled physical velocity component $\bar{\bf v}$ proportional to $\bar\eta\delta$. This allows for a velocity that is a power law in radius and varies on cones and logarithmic spirals. It  makes physical sense to have all of these functions the same on cones and spirals  (near the plane), if all quantities refer to the turbulence generating the dynamo. 

%This might be the case if the gravitational spiral arm potential is responsible  for both the macroscopic and  turbulent velocities.

In any case we proceed by assuming the parameters $\{u,v,w,\Delta\}$ to be independent of $\kappa$. Rather than attempt a general solution for $\bar {\bf A}(\kappa)$, we use the linearity of the equations to make the spiral modal ans\"atz
\be
\bar{\bf A}={\bf C}e^{ip\kappa},~~~~p\equiv \frac{m}{q}.\label{eq:Camp}
\ee
Here $m$ is the mode number and the form of $p$ is required for azimuthal periodicity.

We will only study the similarity class $a=2$ in this work. This implies a global constant with Dimensions equal to that of specific angular momentum (\cite{Hen2015}).  Equations (\ref{eq:Aformsum}), (\ref{eq:vformsum}), (\ref{eq:alphaeff}) and (\ref{eq:diffeff}) consequently indicate the assumed radial dependences. These are, including the modal ans\"atz and remembering equations (\ref{eq:trans1}) and (\ref{eq:logspiral}), 
\bea
{\bf A}(\kappa)&=&{\bf C}\exp{(i\frac{m}{q}\kappa)}\equiv {\bf C}\exp{(i\frac{m}{q}\xi)}\cdot\exp{(-\frac{m~ s1}{q}\frac{z}{r})},\label{eq:Aexp}\\
{\bf v}&=&\bar{\bf v}/(\delta r),\label{eq:vexp}\\
\alpha_d&=&\bar{\alpha_d}/(\delta r),\label{eq:alphaexp}\\
\eta&=&\bar\eta.\label{eq:etaexp}
\eea

Substituting these dependences into the equations for $\bar{\bf A}$ we obtain the homogeneous modal equations 
\bea
0&=&C_R[p^2(1-q^2)-ip(qv+iw ~s1)]+C_\phi[p^2q+ip(v-q-i\Delta~ s1)+v]\nonumber\\
&+&C_Z[ip^2~ s1+ip(w+q\Delta)],\label{eq:rowr}\\
0&=&C_R[p^2q+ip(qu+q+i\Delta~ s1)]-C_\phi[ip(u+iw~ s1)+u+1]\nonumber\\
&+&C_Z[ip^2q ~s1-ip(\Delta-qw)],\label{eq:rowphi}\\
0&=&C_R[ip^2 ~s1-ip(q\Delta-iu ~s1)+C_\phi[iqp^2 ~s1+ip(\Delta+iv~ s1)+\Delta]\nonumber\\
&-&C_Z[p^2(1+q^2)+ip(u+qv)],\label{eq:rowz}
\eea
which are invariant if $z$ and $m$ are negative together. The vector constant ${\bf C}$ may be complex at this stage as it contains both phase and amplitude information.

Setting the determinant of the coefficients of these last equations to zero yields the conditions on the parameters $\{\Delta,u,v,w\}$ for a steady dynamo to exist. An advantage of taking $a=2$ is that only in this case is the divergence of the velocity zero, provided that the set $\{u,v,w\}$ is independent of $\kappa$. This allows the background density to be considered uniform in the steady state if the velocity is non-zero. However we do not solve the dynamical problem in this paper.

%As a sample higher order solution that serves to justify our approximation somewhat, we anticipate the modal solution with ${\bf v}={\bf 0}$ by setting 
%$\Delta=\pm p q=\pm m$ and $a=1$. We remove the $\xi$ dependence from equations (A\ref{eq:AtwovariablefieldR}), (A\ref{eq:Atwovariablefieldphi}) and (A\ref{eq:AtwovariablefieldZ}) by setting $\bar{\bf b}=\tilde{\bf b}e^{(ip\xi)}$. Then these equations yield:
%\bea
%\tilde b_r&=& \pm(i\tilde b_z-\frac{\tilde b'_\phi}{p q}),\nonumber\\
%\tilde b'_\phi&=&\pm\frac{(i-p)\tilde b_\phi}{1\pm i Z},\label{eq:samplesol}\\
%(1\mp i Z)\tilde b'_z\mp p\tilde b_z&=&-i \tilde b_\phi(p q-\frac{(1+ip)}{q(1\pm iZ)^2}).\nonumber
%\eea
%If we work to first order in $Z$, then the  phi component becomes 
%\be
%\tilde b_\phi =C_\phi e^{(\mp p Z)},\label{eq:bphisample}
%\ee
%where $C_\phi$ is a complex constant. Then $\tilde b_z$ and $\tilde b_r$ have the same exponential dependence to lowest order, each multiplied by a linear function in $Z$ that is proportional to $C_\phi$.  There is also a dependence on $p$ and $q$. The linear dependence on $Z$ may be neglected in each case provided that $Z<1$, $q<1$ and $p>1$. Then the $Z$ dependence of this sample solution reduces to  the dependence on $Z$  that appears  uniquely through $\kappa$ in the modal analysis.  The divergence of this solution is only zero to the extent that terms in $Z$ may be neglected and $p^2\gg1$. This consistent with what we assume in our approximation.

We will proceed with the approximate modal analysis of the steady dynamo in  the section below.  In appendix A we show the exact equations and the nature of our approximation. In Appendix B  we introduce the spiral modal analysis of gravitational arms. This analysis is well known \citep{BT2008}, but not as a scale-invariant class.

\section{Steady Dynamo Examples}

$\bullet${An Exact Analytic Example}

We find an exact analytic solution of the scale invariant {\it magnetic field} equations given in Appendix A.  The solution is solenoidal and is easily stated implicitly in its complex form, but the real parts are complicated so that we omit these explicit expressions for brevity . Our main concern is to show that  this exact solution contains similar behaviour  to that which we find below in our approximate treatment with more general parameters.

The solution is found with $a=2$ and the scaled radial velocity $u=-1$. Moreover we anticipate our approximate results by setting $\Delta=pq\equiv \pm m$.  This gives radial infall in a pattern frame according to $v_r=-\bar\eta\delta/r$. We remove the $\xi$ dependence from equations (A\ref{eq:AtwovariablefieldR}), (A\ref{eq:Atwovariablefieldphi}) and (A\ref{eq:AtwovariablefieldZ}) by setting $\bar{\bf b}=\tilde{\bf b}e^{(ip\xi)}$. Then these equations yield:
\bea
\tilde b_r&=&\frac{im\tilde b_z-\tilde b'_\phi}{\Delta},\label{eq:brexact}\\
\tilde b'_\phi&=& \frac{1+ip}{Z\mp i}\tilde b_\phi, \label{eq:bphiexact}\\
0&=&\tilde b'_z(Z\pm i)-ip\tilde b_z-\tilde b_\phi\Delta-\frac{\tilde b''_\phi}{\Delta}.\label{eq:bzexact}
\eea

These equations are linear and are readily solved for the complex forms. Here we consider only the azimuthal equation, for which the exact solution is 
\be
\tilde b_\phi=C_\phi \exp{((1+ip)\ln{(1\pm iZ)}\mp \frac{i\pi}{2})},\label{eq:exactsol}
\ee
where $C_\phi$ is a complex constant. If we treat $Z$ as small,  expand this result, and restore the $\xi$ dependence we obtain
\be
\bar b_\phi\approx C_\phi e^{(\pm i(Z-\frac{\pi}{2})\mp pZ+ip\xi)}.\label{eq:approxbphi}
\ee
Thus  on taking the real part, the exact solution reduces in this limit to  modes  that are either exponentially damped or growing (depending on whether $\Delta$ is positive or negative) with $Z$.  Moreover these  oscillate in $Z$ and $\xi$, which suggests twisted loops near the plane. This is  form that we find  also in our approximate procedure, which  procedure  allows a more general choice of parameters.

%We must find a real root of  equation (\ref{eq:detA}), whose existence we can guarantee by requiring the coefficients of the cubic to be real. This condition is slightly less general than setting the real and imaginary parts of the determinant equal to zero, which we also pursue below. 

%$\bullet$ {Special General Case}

%Requiring real coefficients in the cubic requires that 
%\be
%w=-\frac{m}{q},~~~~u+qv=-\frac{1}{2},\label{eq:woutflow}
%\ee
%and 
%\be
%\Delta^2=\frac{1}{4}+(\frac{m^2}{q^2}-v^2)(1+q^2),\label{eq:deltacond}
%\ee
%and the right hand side must be positive. The sign of $\Delta$ may however be positive or negative, corresponding to the sign of $\bar\alpha_d$.
 %These conditions for trailing spirals ($q>0$) imply vertical and radial infall (unless the pattern speed azimuthal velocity $v<0$) for $m>0$. When $m<0$ vertical outflow is possible with inward radial velocity and trailing arms. The scaled velocities have all been taken constant in $\xi$. We note that the $m<0$ mode gives outflow above the plane ($z>0$) for trailing arms, while the $m>0$ mode gives outflow below the plane. The sign of the mode number is opposite above and below the plane for inflow. In general we find that the change in sign of the mode number allows us to continue  corresponding solutions across the plane.
 
$\bullet $ {The general  Approximate Case}

We have to solve the homogeneous equations (\ref{eq:rowr}). (\ref{eq:rowphi}) and (\ref{eq:rowz}) for the amplitudes of the dynamo vector potential. In fact we can only find two ratios, which leaves us with the expected arbitrary constant. This is only possible if the determinant of the coefficients of these equations is zero, which gives us the restriction on our parameters for a steady state to exist. The determinant condition takes the form of a cubic for $\Delta$ with complex coefficients depending on the velocity components, namely 
\bea
&&\Delta^3+(-im+uq+2mws1-2imu+qw^2+qv^2+u^2q-2imqv-m^2q)\Delta/q\nonumber\\
&+&(-mvs1+im^2-qvw+imqw)/q=0.\label{eq:detA}
\eea
 We recall that $\alpha_d$ and $\bar\eta$ may be the same arbitrary function of $\kappa$, given this dispersion relation. From equations (\ref{eq:vformsum}) and (\ref{eq:scaledvel}) we see that the physical velocity may also be the same function of $\kappa$ multiplied by $1/r$, provided that the scaled velocity components ${\bf u}$ are constant.

The general case is found by setting the real and imaginary parts of equation (\ref{eq:detA}) equal to zero and requiring a real solution. These conditions become 
\bea
qw+ms1&=&\Delta(1+2u+2qv)\nonumber\\
 0&=& \Delta^3+(u^2+v^2+w^2+u-m^2+\frac{2mws1}{q})\Delta-\frac{v}{q}(qw+ms1).\label{eq:gencons}
\eea
Combining these two equations yields finally another equation for $\Delta$ in the form
\be
q^2\Delta^2=(1+q^2)(m^2+v^2)-(ms1+qw)^2-(v-qu)^2+q(v-qu),\label{eq:gendelta}
\ee
which must hold in order that the  $\Delta$ from first of equations (\ref{eq:gencons}) be a root of the second of these equations. It is not necessarily a root of the equation (\ref{eq:gendelta}) however.  For that to be true there is a restriction on the velocities found by inserting the value of $\Delta$ from equation (\ref{eq:gencons}) into equation (\ref{eq:gendelta}), which takes the form   
\be
(ps1+w)^2\left(1+\frac{q^2}{(1+2u+2qv)^2}\right)=(p^2+\frac{v^2}{q^2})(1+q^2)-(\frac{v}{q}-u)^2+(\frac{v}{q}-u).\label{eq:genvelcon}
\ee
This consistency condition must be checked in each special case below. One  has also to insist on an integer mode number.

We note that the combination~ $qu-v$~ is proportional to the velocity parallel to the  magnetic spiral arm in the plane ($u+qv$ is proportional to the velocity perpendicular to the  magnetic arm). Both signs of the square root are possible. Rather than pursue the solution for the coefficients in the general case, we look at some simpler limits in this work. One should recall at this point that the velocities are in a pattern frame of the magnetic structure. Only if we place ourselves in the systemic frame will the azimuthal velocity correspond to the galactic rotation velocity.

$\bullet$ {All velocities are zero or parallel: Case A}.

Here either ${\bf v}$ is zero in the pattern frame or ${\bf v}\parallel {\bf B}$. 
Equation (\ref{eq:gendelta}) gives immediately that $\Delta^2=(ms1)^2$ (as in our sample analytic solution), but the first of equation (\ref{eq:gencons}) shows that we must extract the positive root  here so that 
\be
\Delta=ms1.\label{eq:deltazero} 
\ee
Hence a steady, stationary (zero velocity in the pattern frame),~ $\alpha_d^2$~ dynamo requires the turbulent (helical) Reynolds number to be equal to plus or minus the mode order. This corresponds to magnetic field that is either decreasing or increasing with $z$ respectively.  A mode determination would thus give directly the value of this quantity,  which can be expected to be of order unity.  

The  helicity  $\alpha_d$ will change sign across the disc with the mode number $ms1$, if a direction is assigned to this quantity by (say) a right-hand rule relative to the positive direction of $z$ . A change in sign then yields the same handedness  relative to the rising direction on crossing the galactic plane. We assume a positive diffusion coefficient.

%We remark that this case also corresponds to a configuration where ${\bf v}\parallel {\bf B}$, since then ${\bf v}$ will not appear in the dynamo equation (\ref{eq:Afield}). This allows us to make contact with the problem of the lagging halo, as discussed in \citet{HI2016}.

$\bullet$ {Only an azimuthal velocity is present: Case B}

This case allows for an $\alpha/\omega$ dynamo in the pattern frame of reference.
Formally equations (\ref{eq:gencons}) and (\ref{eq:gendelta}) give 
\bea
&\Delta&=\frac{ms1}{1+2qv},\nonumber\\
&\Delta^2&=m^2+v^2+\frac{v}{q}\label{eq:alphaomega}
\eea
However reconciling the two values of $\Delta$ is found to require ($v\ne 0,\ne-1/q$) 

\be
\left(\frac{4q^2m^2}{(1+2qv)^2}\right)=-1,
\ee
which is not possible for real $m$,$q$,$v$. We conclude that there is no steady pure $\alpha/\omega$ dynamo (identifying $v$ with the background disc velocity in the pattern frame), unless 
\be
v=-\frac{1}{q}.~~~~~\Delta=-ms1. \label{eq:alphaomegavcon}
\ee

The helicity will again change sign across the plane with the change in  $ms1$. 
%This implies that the arms should be leading the azimuthal velocity (at least in the pattern frame), since this is clearly so when $q<0$, and if $q>0$ as for trailing arms then  $v$ becomes retrograde ($<0$) in the pattern frame.

$\bullet$ {Only a vertical velocity is present: Case C}

This case has some considerable interest since a vertical velocity is often required to quench the evolution of a turbulent dynamo.

Conditions (\ref{eq:gencons}) and (\ref{eq:gendelta}) become 
\bea
\Delta&=&qw+ms1,\nonumber\\
\Delta^2&=&m^2-2\frac{m}{q}ws1-w^2.\label{eq:outflowonly}
\eea
%When $\Delta=m$, $w=0$ and we reduce to the zero velocity case. Hence the new feature is when $\Delta=-m$ and $w=-2m/q$. This requires infall for trailing arms ($q>0$)when $m>0$ or outflow when $m<0$ above the plane. Once again we change the sign of the mode number to extend a solution below the plane.  A negative sign of $\Delta$ refers to the sign of the helical turbulence through ~$\bar\alpha_d$. Once again the mode number yields the Reynolds number $\Delta$, while the sign depends on inflow or outflow.  
Eliminating $\Delta$ between these two expressions gives ($w\ne 0$)
\be
w=-\frac{2ms1}{q},\label{eq:wonly}
\ee
so that the ratio of alpha velocity to diffusion velocity is again 
\be
\Delta=-ms1.\label{eq:wDelta}
\ee

One can change from infall  to outflow by taking $ms1<0$ above the galactic plane. However  then we find  the strength of the magnetic field increasing   with positive $z$. This is effected below the plane by taking $ms1>0$.   Antisymmetry for the tangential magnetic fields  applies to both infall and outflow (see below) on crossing the plane. The vertical velocity  is scaled by the diffusion velocity $\eta\delta$, which has a certain arbitrariness due to the choice of turbulent scale $1/\delta$.

$\bullet$ {Only a radial velocity is present: Case D}

In this case we have from equations (\ref{eq:gencons}) and (\ref{eq:genvelcon}) that  ($u\ne -1$)
\be
\Delta=\pm\frac{1}{2},~~~~~u=\pm m-\frac{1}{2}\label{eq:onlyucon}
\ee
Therefore we have only a very restricted steady dynamo wherein $\Delta=\pm 1/2$ and the radial velocity is slower  when directed radially outward.
The excluded case $u=-1$ satisfies all of the requirements with $\Delta=-ms1$.  It corresponds to our exact example.

%$\bullet$ {The velocity perpendicular to the arm is zero: Case E}

%Here we require that $u=-qv$, whereupon equations (\ref{eq:gencons}) together imply that 
%\bea
%\Delta&=& qw+ms1,\nonumber\\
%\Delta^2&=& m^2+qv(1-qv),
%\eea
%and eliminating $\Delta$ between these last two equations gives
%\be
%qv(1-qv)=(qw+ms1)^2-m^2.\label{eq:perpzero}
%\ee
%Hence if $w>0$ (outflow at positive $z$), we must have $v<1/q$, and if $w<0$ (inflow) we must have $v>1/q$  assuming $ms1>0$. If $ms1<0$, (growing field in $z$) the conditions for inflow  or outflow on $v$ are interchanged. On crossing the plane $w<0$ corresponds to outflow so that $ms1<0$ gives $v<1/q$ and thus continuity of rotation across the plane. Inflow requires again $v>1/q$.   If below the plane $ms1>0$ (growing field in $-z$), one requires $v>1/q$ for outflow and $v<1/q$ for inflow. This assumes $v>0$. If $v<0$, $ms1$ and $w$ must always have opposite signs, which for outflow means magnetic fields growing away from the plane.    

%$\bullet$ {The velocity parallel to the arm is zero: Case F}

%In this case $ v=qu$, and hence from equations (\ref{eq:gencons}) there follows 
%\bea
%\Delta&=&\frac{qw+ms1}{1+2u(1+q^2)},\nonumber\\
%\Delta^2&=& \frac{(m^2+q^2u^2)(1+q^2)}{q^2+(1+2u(1+q^2))^2}.
%\eea
 %Together these require 
%\be
%(qw+ms1)^2\left(\frac{q^2}{(1+2u(1+q^2))^2}+1\right)=(m^2+q^2u^2)(1+q^2),\label{eq:zeroparallel}
%\ee
%which is possible for a variety of choices of $u$ and $w$. 

We turn in the next section to the form of the magnetic field in terms of the complex coefficients ${\bf C}$. In this section we have found general conditions on the parameter $\Delta$ and the scaled velocities under which a steady dynamo (`kinematic', because the velocities are used as parameters) can exist.

\section{Magnetic Field Structure }

We calculate the magnetic field from equation (\ref{eq:Aexp}) and ${\bf B}=\nabla\wedge {\bf A}$. We take the real part. We usually take the arbitrary constant to be real and equal to $C_\phi$. In that case we find the field (divided by $\sqrt{4\pi\rho}$ for arbitrary density $\rho$) in the form 
\bea
b_r&=&\!\!C_\phi\frac{p}{r}\exp{(-\frac{p z s1}{r})}[(s1-q\frac{Im(C_z)}{C_\phi})\cos{(p~\xi)}-q\frac{Re(C_z)}{C_\phi}\sin{(p~\xi)}],\label{eq:bR}\\
b_\phi&=& \!\!C_\phi\frac{p}{r}\exp{(-\frac{p z s1}{r})}[(\frac{Im(C_z)}{C_\phi}-\frac{Re(C_r)}{C_\phi}s1-\frac{z}{r}\frac{Re(C_z)}{C_\phi}s1)\cos{(p~\xi)} \nonumber\\
&+&\!\!(\frac{Im(C_r)}{C_\phi}s1+\frac{Re(C_z)}{C_\phi}+\frac{Im(C_z)}{C_\phi}s1\frac{z}{r})\sin{(p~\xi)}],\label{eq:bphi}\\
b_z&=&\!\!\frac{C_\phi}{r}\exp{(-\frac{p z s1}{r})}[(1+qp\frac{Im(C_r)}{C_\phi}+\frac{p z s1}{r})\cos{(p~\xi)} \nonumber\\
&+&(qp\frac{Re(C_r)}{C_\phi}-p)\sin{(p~\xi)}].\label{eq:bz}
\eea 

The real and imaginary parts of the coefficients ${\bf C}$ are indicated explicitly. It is readily shown numerically that this magnetic field is divergence free to one part in $10^{10}$  for any particular set of parameters and any particular location in space.  The behaviour  crossing the galactic disc  depends on the behaviour of the coefficients at this transition. In many cases however we find that the tangential magnetic field changes sign with $ms1$ and $z$ changing together, while the vertical field remains continuous. This changes the handedness of the magnetic field across the plane.

\subsection{Zero Velocity Steady Dynamo Field Structure: Case A}

This is the pure $\alpha^2$ dynamo with diffusion. It stands in contrast to the $\alpha/\omega$ dynamo with diffusion that arises with non-zero ${\bf v}\wedge {\bf b}$.

In the  general modal analysis, the exponential decline (or growth)  from $z=0$ is only one way in which the disc is present. There is also the question of boundary conditions. Two possibilities arise. In one case all components of the magnetic field are considered to be continuous. The magnetic field at $z<0$ is then identical to that at $z>0$. This is the symmetric boundary condition. We do not find this possibility in this  pure $\alpha^2$ example. 

In the second case the vertical field is continuous but the tangential magnetic field is anti-symmetric across the disc such that ${\bf b}^-_{\parallel}=-{\bf b}^+_{\parallel}$ at $z=0$. In order to arrange this with spiral modes we find that one must hold the switch  $s1$ constant at $+1$ while changing the sign of $z$ and $m$ (hence also $\Delta$). This is because the determinant that yields the coefficients $C_r$ and $C_z$ when ${\bf v}\parallel {\bf b}$ (or zero) is invariant when crossing the plane under this condition. The vertical component of the field is  properly continuous, but its derivative changes sign according as the plane is approached from `above' or `below' the plane. In an average sense  $(db_z/dz)_0=0$, which one expects with this boundary condition on averaging the continuity equation. 

A closed magnetic field loop can be constructed for the anti-symmetric boundary condition by calculating the arc above the plane  from a first crossing of the galactic plane to a second crossing. Then one changes the symmetry of the field (which reverses the direction of increasing arc length) and computes the arc below the plane that returns to the first crossing of the plane. We illustrate this in the figures. 

% Alternately, one can introduce small errors at the plane crossing by allowing the change in field symmetry to only roughly (to within a step length) coincide with a plane crossing. This converts the perfect loops into a helical structure wrapped around the magnetic arms. Such errors may correspond to magnetic field `sources' (stellar fields) in the disc. 

\begin{figure}%[p]
\begin{tabular}{cc} %This will make a two-column figure
\rotatebox{0}{\scalebox{0.65} %change the angle and scale as you need
{\includegraphics{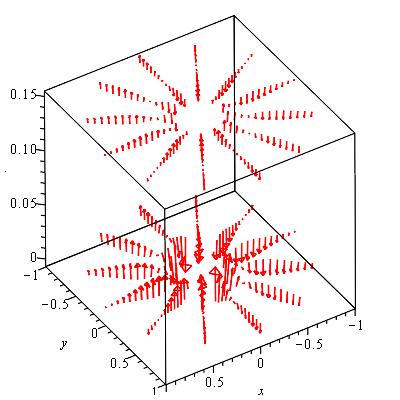}}}&
\rotatebox{0}{\scalebox{0.65} %change the angle and scale as you need
{\includegraphics{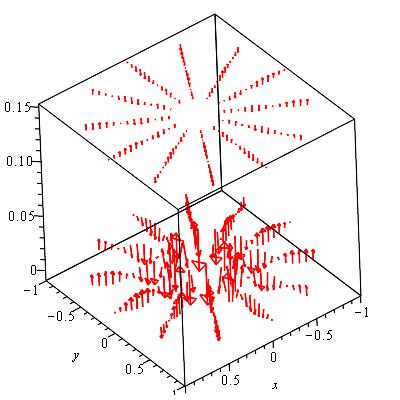}}}\\
{\rotatebox{0}{\scalebox{0.4} %change the angle and scale as you need
{\includegraphics{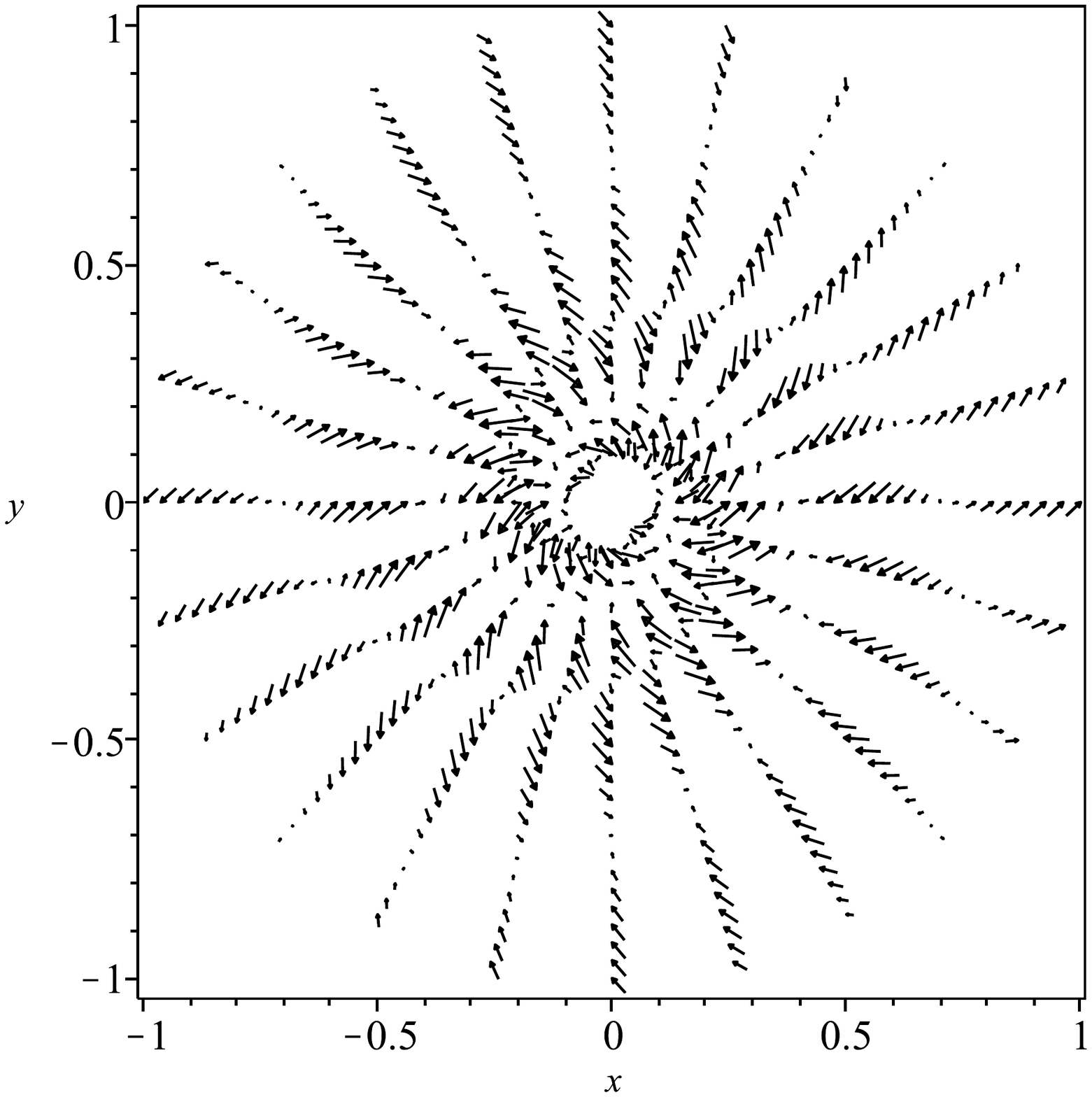}}}}&
\rotatebox{0}{\scalebox{0.4} %change the angle and scale as you need
{\includegraphics{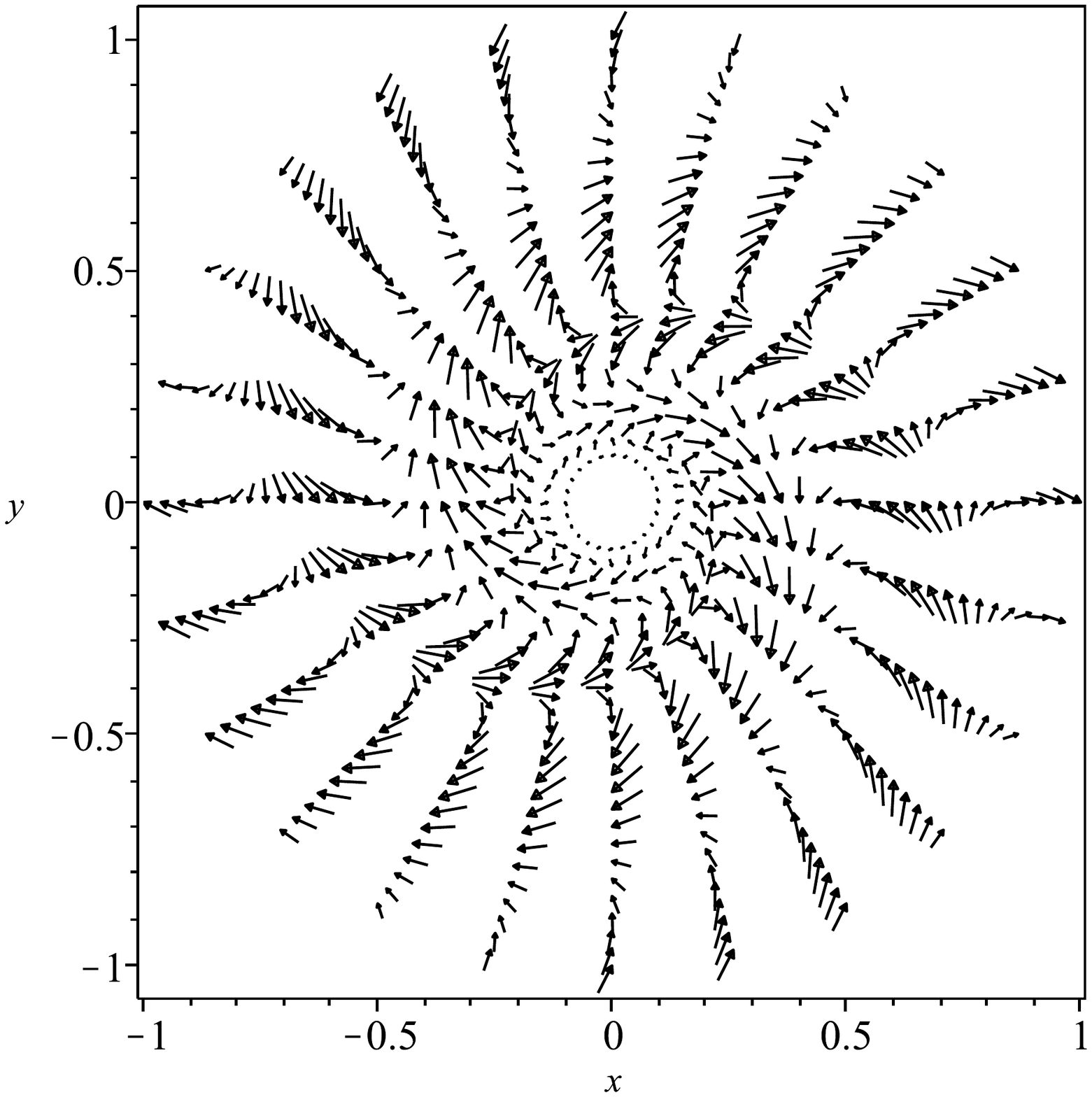}}}
\end{tabular}
\caption{At upper left we show a vector plot of the magnetic field near the plane for $z\ge 0$. The switch $s1=+1$. The mode is $m=1$ and the tangent of the pitch angle is again $q=0.4$ ($\approx 22^\circ$). On the right is the same case for the $m=2$ mode. The ratio $z/r$ is never greater than $0.6$ in these figures, which supports our approximation over most of the plot.  The region $r<0.25$ is excluded and the radius of the disc is $r=1$. The two dimensional plot at bottom left is for $m=2$ and $s1=1$ and at bottom right $m=-2$ and $s1=-1$ so that $\Delta=2$ in each csse.  The pitch angle is $\arctan{(0.4)}$  and the cut is at $z=0.05$ . The arrows are anchored by the head.}    
\label{fig:vzeroField3d}
\end{figure}

Having confirmed that the determinant of the coefficients is zero, equations (\ref{eq:rowr}), (\ref{eq:rowphi}) and (\ref{eq:rowz}) are readily solved for $C_r$ and $C_z$ as 
\bea
C_r&=& -\frac{i}{m}C_\phi,\label{eq:Crzero}\\
C_z&=&s1\frac{2iqm+1}{m(1+q^2)}C_\phi\label{eq:Czzero}
\eea
By taking the real and imaginary parts of these expressions, we determine from equations (\ref{eq:bR}), (\ref{eq:bphi}) and (\ref{eq:bz}) the magnetic field at any point to within an arbitrary multiplicative quantity $\sqrt{4\pi\rho}$. We recall that $C_\phi$ is an amplitude of the  azimuthal vector potential and has the Dimensions of this potential. Figure (\ref{fig:vzeroField3d}) shows some aspects of the magnetic field structure in the galactic disc at $z=0^+$.

\begin{figure}%[p]
\begin{tabular}{cc} %This will make a two-column figure
\rotatebox{0}{\scalebox{0.4} %change the angle and scale as you need
{\includegraphics{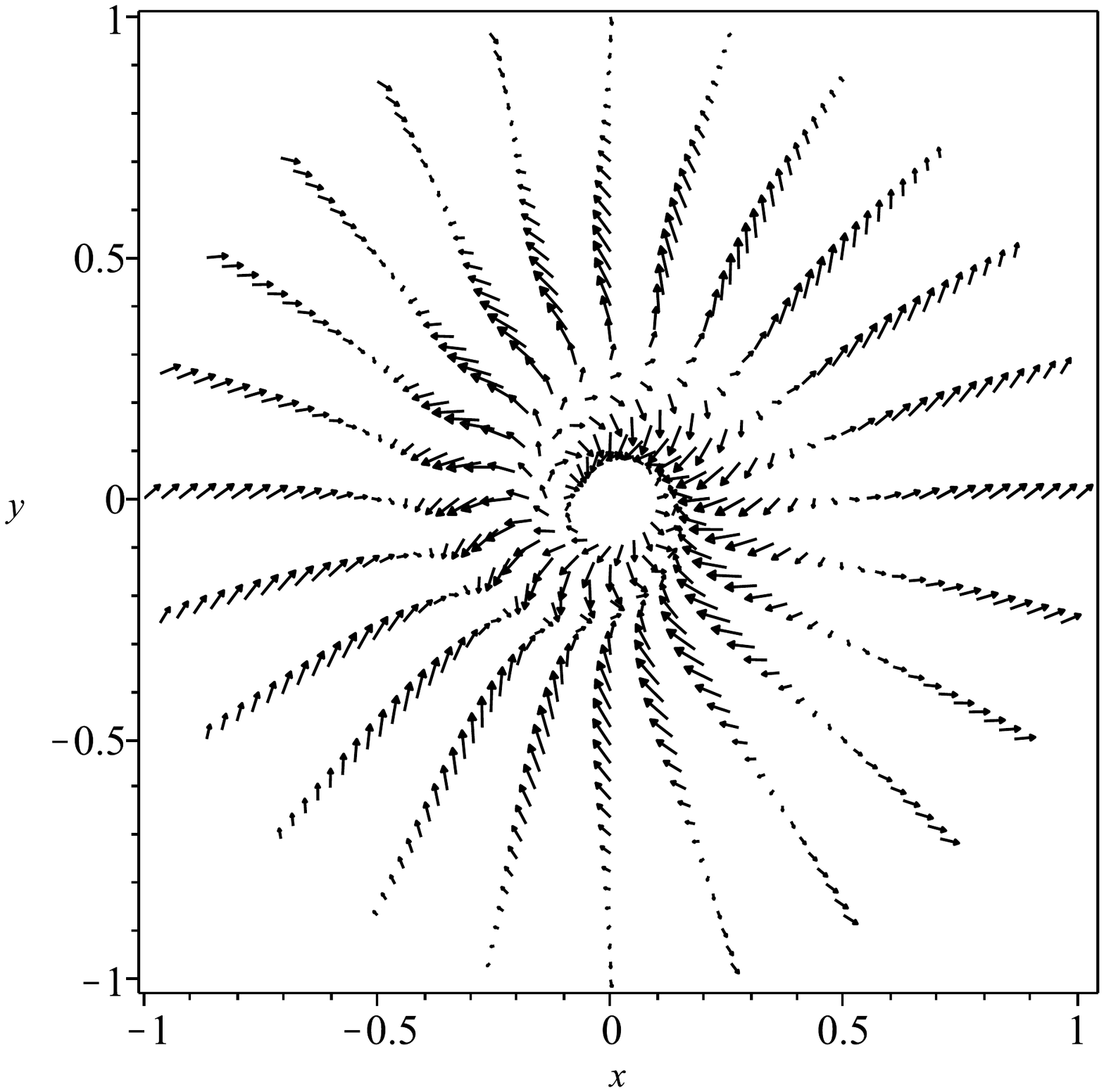}}}&
\rotatebox{0}{\scalebox{0.4} %change the angle and scale as you need
{\includegraphics{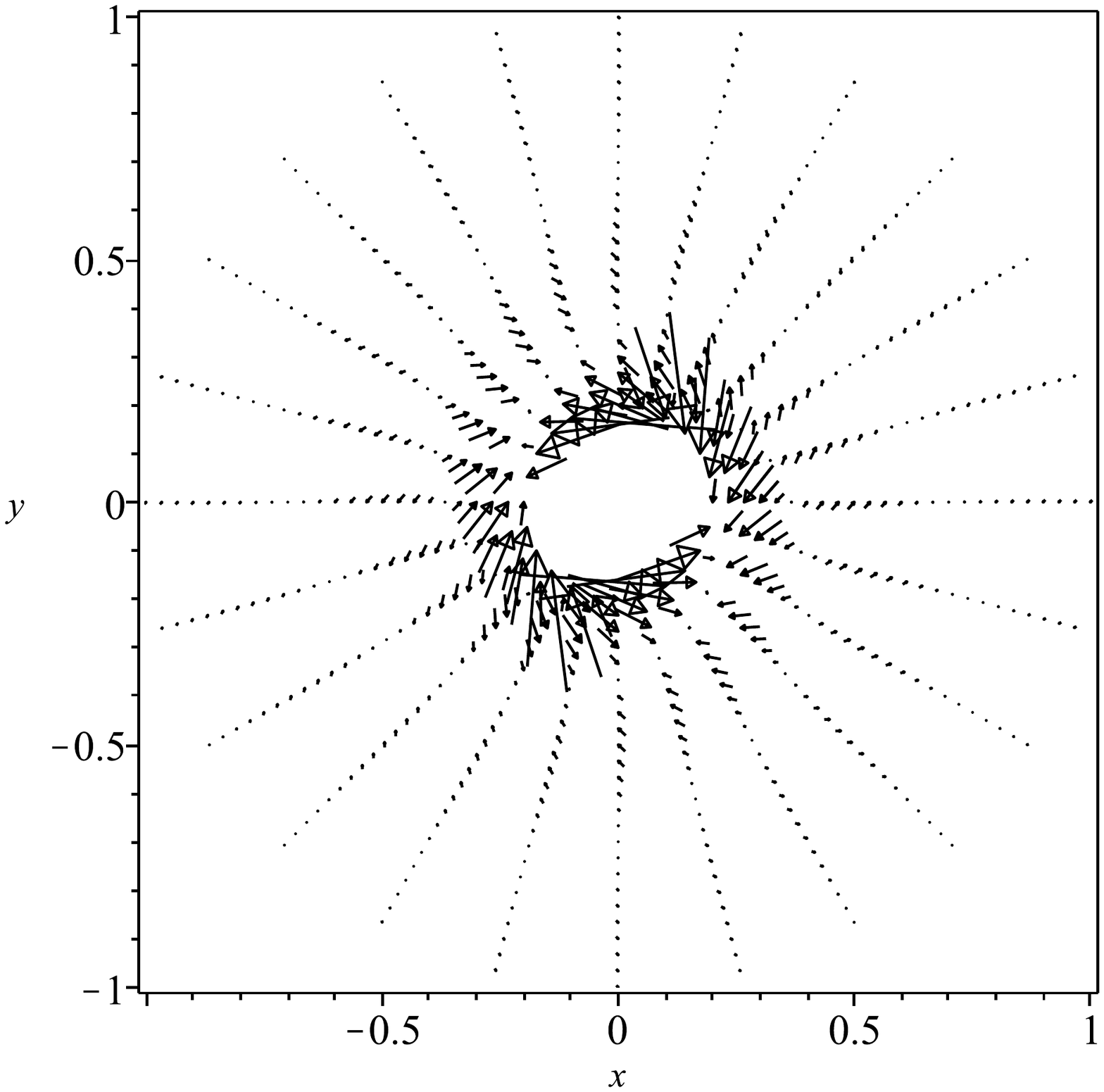}}}\\
{\rotatebox{0}{\scalebox{0.4} %change the angle and scale as you need
{\includegraphics{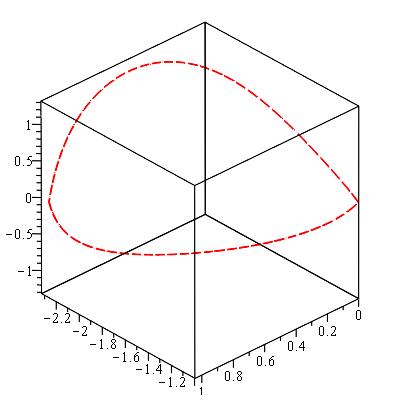}}}}&
\rotatebox{0}{\scalebox{0.4} %change the angle and scale as you need
{\includegraphics{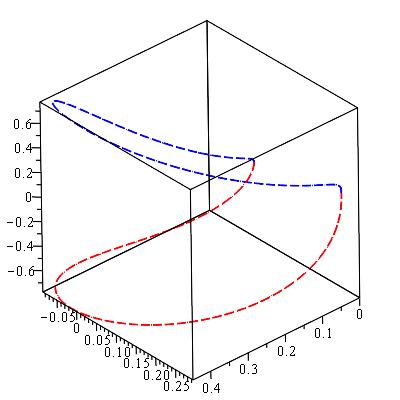}}}
\end{tabular}
\caption{The figure at upper left shows the magnetic vectors in a horizontal cut through the magnetic halo at $z=0.1$, when $q=0.4$, $m=1$, $s1=1$ and $z=0.1$. Radii $r<0.1$ are excluded  and the disc radius is $r=1$. At upper right the magnetic vectors in the same cut are shown for the mode $m=-2$ , $s1=1$ and otherwise for the same parameters as at upper left. At lower left a magnetic field loop that begins in the plane at $\{r,\phi,z\}=\{1,-\pi/2,0.00001\}$ is shown. Once again $m=1$ and $q=0.4$ above the plane but we change the sign of $m$ below the plane to change the handedness of the loop. At lower right the loop begins at $\{r,\phi,z\}=\{0.25,\pi/2,0.00001\}$ for the same $m$ and $q$ with the same anti-symmetric boundary condition across the plane. }
\label{fig:vzerodynamo}
\end{figure}

In figure (\ref{fig:vzeroField3d}) we show vectors near  but above the plane, for $m=1$ on the left and $m=2$ on the right. In each case $q=0.4$. We see that the fields  vary rapidly in $z$, radius and azimuth. The variations in radius and azimuth  are correlated with the location of the arms, as may be seen by examining the arms in figure (\ref{fig:vzerodynamo}). There is a suggestion that the field must connect in loops over the arms, but the tangential field is concealed in the figure by the stronger vertical component.  Nevertheless, a careful inspection of the figure does show rapid variation in the sign of the field vector associated with loops near the plane. We illustrate these loops in figure (\ref{fig:vzerodynamo}). 

 We observe that the field strength falls off faster in $z$ at smaller $r$ as expected from the scale height in the self-similar form.   The $m=2$ mode falls off generally more quickly in $z$ than does the $m=1$ mode. The extent to which spiral structure is `lifted' into the halo is important for the rotation   (and polarization) measures of edge-on galaxies. Normally the complete field will include an axially symmetric ($m=0$) component, which has been treated  in a companion paper.

The bottom left image in this figure shows a two-armed spiral with positive helicity ($m=\Delta>0$) at $z>0$. This produces  magnetic  `polarization arms'  wherein the magnetic vector is at a considerable angle to the spiral arm axis. The figure at bottom right differs from bottom left only in having a negative mode number. The arms are still mainly polarization arms at large radius.

At upper left in figure (\ref{fig:vzerodynamo}) we show a cut through the magnetic halo at $z=0.1$, as produced by the $m=1$ dynamo. The pitch angle is $22^\circ$.  At upper right the image shows the same cut for $m=-2$, $s1=+1$, and the same pitch angle.  This illustrates a case where the magnetic field is increasing with increasing $z$, most rapidly at small $r$. 

In all cases the strongest magnetic structures are not magnetic arms in the sense that the magnetic field lies along the arm. They are rather `polarization' arms where the linear polarization (for a face-on galactic view) would be strong in a spiral pattern. Since some of the vectors must lie on loops above the plane, we can also expect variations in Rotation Measure (RM) over the plane.  The loops calculated in the lower part of the figure extend over large regions of the plane and over the arms. Such loops manifest zero divergence.  There should be RM variations  on the same scale.

%We display the modes $m=-1$ and $m=-2$ preferentially in the upper part of the figure since they produce magnetic spiral arms with the magnetic field directed along the spiral. The positive modes are similar but the magnetic vectors are not aligned with the arms. This may be seen in the image at lower left of the figure where the $m=1$ mode is shown for the same parameters. Hence there are only `polarization arms' in the positive modes  at $z=0^+$ where the polarization is strong but the vectors are more nearly transverse and do not connect along the arm . The behaviour is reversed below the plane.  Scale invariance ensures that any section of the plane will look similar to these sections, although the absolute amplitudes vary proportional to $1/r$. 

% At lower right the image is of the sum of five modes from $m=-1$ to $m=-5$ for $z=0^+$, all with pitch angle $q=0.4$. The sum approximates to a delta function in $\xi$ as the number of modes increases,  because the amplitude is the same for all of the individual modes. This image reveals an inner magnetic spiral arm (with aligned vectors) which is weaker than a `polarization arm' (unaligned vectors) that lies outside it. The polarization arm appears more `turbulent'. If that arm were to be identified arbitrarily with the stellar /gas arm then the  true magnetic spiral would lie inside this arm with the same pitch angle. The strong field in the normal spiral arm is likely to be depolarized. Adding more such modes (e.g. to $m=-10$) does not change this behaviour very much. 
The field line loops shown in the lower part of figure (\ref{fig:vzerodynamo}) are calculated according to 
\bea
\frac{dr}{ds}&=&\frac{\frac{B_r}{|B_r|}}{\sqrt{1+\frac{B_\phi^2+B_z^2}{B_r^2}}},\label{eq:drds}\\
\frac{d\phi}{ds}&=& \frac{1}{r}\frac{B_\phi}{B_r}\frac{dr}{ds},\label{eq:dphids}\\
\frac{dz}{ds}&=& \frac{B_z}{B_r}\frac{dr}{ds},\label{eq:dzds}
\eea
where $ds$ is along the field line. The sense of increasing arc length ($ds>0$) will coincide with the increasing direction of the tangential coordinates,  when the tangential field components are positive.  These loops must be regarded only as illustrative since the full magnetic field comprises an infinity of such loops. Their existence seems certain, but because of our approximation we can not really follow them to heights greater than a kiloparsec or so above the plane. The loops shown in the figure are pushing the approximation. Nevertheless this structure seems to be characteristic of a pure `alpha squared'  turbulent dynamo as discussed in this section.  The topology  has observational consequences.  

\begin{figure}%[p]
\begin{tabular}{cc} %This will make a one-column figure
\rotatebox{0}{\scalebox{0.6} %change the angle and scale as you need
{\includegraphics{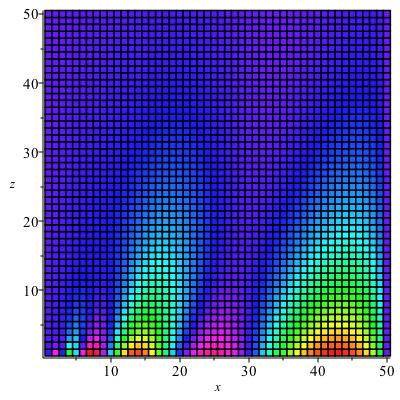}}}\\
\rotatebox{0}{\scalebox{0.6} %change the angle and scale as you need
{\includegraphics{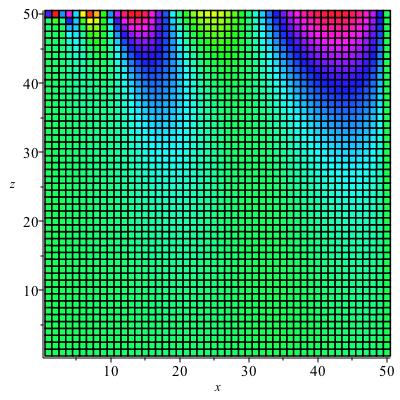}}}
%{\rotatebox{0}{\scalebox{0.5} %change the angle and scale as you need
%{\includegraphics{onehundredgravmodes.eps}}}}
\end{tabular}
\caption{ The figure shows the Faraday screen RM in the first and fourth quadrants for the $m=2$ mode, $q=0.4$ and $s1=1$.  The electron density near the disc is taken constant. The antisymmetry across the disc is apparent, which requires changing the sign of $m$ below the plane. The second and third quadrants would appear similarly, but for an interchange of colours.  They are found by rotating the figure about the left-hand edge (the galactic axis)  according to the left-hand rule and interchanging colours. Positive RM is generally red and blue, while negative RM  is yellow, orange and green. }    
\label{fig:naiveRM}
\end{figure}

 Additional observational consequences follow especially for edge-on galaxies (e.g. \citet{CMP2016}; \citet{SPK2016}). We take a simple model of a cylindrical halo that has the same radius as the galactic disc. We assume the inclination of the galaxy to be $90^\circ$ and calculate (using the dynamo magnetic field) 
 \be
 RM=\int~n_eb_\parallel~ds,\label{eq:RMint}
 \ee
 along the line of sight, distributed over the halo. In order to isolate the effect of the dynamo magnetic field, we take the electron density to be constant. This is easily changed in order to deal with observed cases. The radial variation of the electron density is probably the most important additional parameter, because we are constrained to be near to the plane. We show the first and second quadrants  (the galactic minor axis is the ordinate) in figure (\ref{fig:naiveRM}) for the $m=2$ mode with the pitch angle of $22^\circ$. 
 
 This quantity that we call rotation measure (RM) is only truly relevant to a Faraday screen with no internal sources.   This is not the quantity that is found by the application of rotation measure synthesis. One should ultimately integrate the equation of radiative transfer through the halo and calculate $1/2 arctan(U/Q)$ ($U$,$Q$ are the Stokes parameters) for the exiting radiation. For an optically thin halo this is quite feasible, but we must leave it to a more observationally oriented work.  Small pitch angles approach axial symmetry, and in that case the RM may be considered as that produced on emission at the tangent point to a true magnetic spiral arm.
 
 We see the anti-symmetry across the disc in figure (\ref{fig:naiveRM}) that is characteristic of these  `zero velocity' modes. The $m=1$ mode has fewer reversals, but these extend to larger $z$ so we have not shown them here in view of our approximation. 
 
 The second and third quadrants are readily generated by rotation about the galactic axis (on the left edge and upward) using the left-hand rule. Subsequently the colours must be interchanged. This produces diagonal symmetry over the whole halo. The positive or negative sense of the colours can also be interchanged by changing the sign of the amplitude constant $C_\phi$.
 
  In figure (\ref{fig:naiveRM}) the abscissa gives $r=grid number/50$, so that  one Unit radius is the radial extent of the disc. The ordinate shows $z=\pm grid number/100$, so that the halo extends to only half the disc radius and the structure lies within our approximate region. 
The absolute magnitude of this Faraday screen RM requires fitting a multiplicative constant to the data. However the {\it contrast} in amplitude of peaks of  different signs varies  along the disc in the range from $\sim 2.3$ to $\sim 4$.  Corresponding variations in the polarized intensity in the halo should also occur, with peaks anti correlated with theRM peaks.  Observing such variations can vastly improve our knowledge of galactic halo magnetic fields.

In cases where $ms1<0$ at positive $z$, the contrast in RM between the arms can be very great, as seen at upper right in figure (\ref{fig:vzerodynamo}). 
 Detectable RM oscillations may only be visible at large radius.

\newpage
\subsection{Dynamo with Azimuthal Velocity only: Case B}
This case comprises a simple $\alpha-\omega$ mean field dynamo. It is particularly simple to find the complex coefficients that determine the mean magnetic field. These are  
\bea
C_r&=&\frac{2is1}{q}C_\phi,\label{eq:Crvonly}\\
C_z&=&(\frac{2}{q}-\frac{i}{m})C_\phi.\label{eq:Czvonly}
\eea
 The calculations reveal that it is with $m<0$ and $s1<0$ that true magnetic arms are produced. The positive mode number and positive $s1$ give only polarization arms. There is a case with the magnetic field increasing with $z$ wherein $m>0$ and $s1<0$. Should $m=1$ one obtains very strong true magnetic arms only at small radius. The contrast in the RM oscillation amplitude that results is very large, so that one sign is likely to be undetectable. The $m=2$, $s1<0$ case gives only strong polarization arms.
 
 The RM produced by the mean field when $m<0$ and $s1<0$ is rather similar to forms found below in case C. In particular the $m=-2$, $s1=-1$  mode  RM is almost identical to that of figure (\ref{fig:RMplotW-2-1}) regarding oscillation on the same side of the  galactic minor axis, so that we do not repeat it here.
 
\begin{figure}%[p]
\begin{tabular}{cc} %This will make a one-column figure
\rotatebox{0}{\scalebox{0.6} %change the angle and scale as you need
{\includegraphics{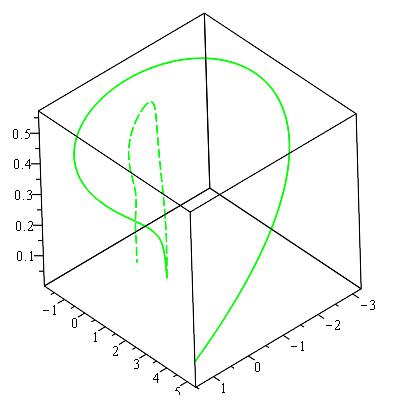}}}
%\rotatebox{0}{\scalebox{0.6} %change the angle and scale as you need
%{\includegraphics{zerovelRMm22nd.jpg}}}
%{\rotatebox{0}{\scalebox{0.5} %change the angle and scale as you need
%{\includegraphics{onehundredgravmodes.eps}}}}
\end{tabular}
\caption{ The solid curve shows a loop above the galactic disc calculated for the $m=-1$, $s1=-1$ mode. The dashed curve is for the $m=+1$, $s1=-1$ mode. Both curves are started at $r=0.75$, $\phi=3\pi/4$ and $z=0.0015$. The value  $q=0.4$ for both loops. }    
\label{fig:vonlyloops}
\end{figure}

 Moreover the magnetic field above  the disc forms high loops in our calculations. Figure (\ref{fig:vonlyloops}) shows a loop extending to large radius (solid line) for the $m=-1$, $s1=-1$ mode, plus a smaller loop (dashed line) present in the $m=+1$, $s1=-1$ mode. The latter field increases with increasing $z$. Both field lines start at the disc at the same azimuth and radius, but the dashed loop rather quickly moves to the centre of the galaxy, where it crosses the plane. We do not pursue this case further here, but when fitting data it should be combined with case C  as studied below.
 
As a corollary to this section, the assignment $v=-1/q$ also allows a dynamo that includes vertical outflow. The parameters are :
\be
~~~~~~~~~~~~~~~~u=0,~~~~v=-\frac{1}{q},~~~~w=-\frac{2ms1}{q},~~~~\Delta=ms1.\label{eq:vwdynamo}
\ee
The coefficients required in order to calculate the magnetic field are
\bea
C_r&=&-i\frac{(2ip-2p^2+1)}{pq(1+2ip)}C_\phi,\label{eq:Crvw}\\
C_z&=&\frac{2(i-p)}{q(1+2ip)}C_\phi.\label{eq:Czvw}
\eea
This case is likely to be prominent when fitting observed data, but in the interests of brevity we do not study it in this introductory paper.
\newpage
\subsection{Dynamo with Velocity Perpendicular to the Galactic Plane: Case C}

This case includes the effect of gas either falling onto or ejected from the galactic plane  (above or below the plane), according to $w=-2ms1/q$ and $\Delta=-ms1$. That is, taking $ms1<0$ , $q>0$ and $z>0$  allows gas to be flowing out of the plane, while $ms1>0$ describes accretion onto the plane with. Below the plane we must have $ms1>0$ for outflow and the reverse for inflow.  

Once again, having first set the determinant of the coefficients equal to zero, equations (\ref{eq:rowr}), (\ref{eq:rowphi}) and (\ref{eq:rowz}) are readily solved for $C_r$ and $C_z$ in the form
\bea
C_r&=& \frac{2m^2+q^2}{mq(2m+iq)}C_\phi,\label{eq:Crwonly}\\
C_z&=&s1C_\phi\left(\frac{2im^2(1+q^2)+iq^2-2mq^3}{mq(1+q^2)(2m+iq)}\right).\label{eq:Czwonly}
\eea
We are now able to find the magnetic field by taking the real and imaginary parts of these quantities and substituting these into equations (\ref{eq:bR}), (\ref{eq:bphi}) and (\ref{eq:bz}).

\begin{figure}%[p]
\begin{tabular}{cc} %This will make a two-column figure
\rotatebox{0}{\scalebox{0.4} %change the angle and scale as you need
{\includegraphics{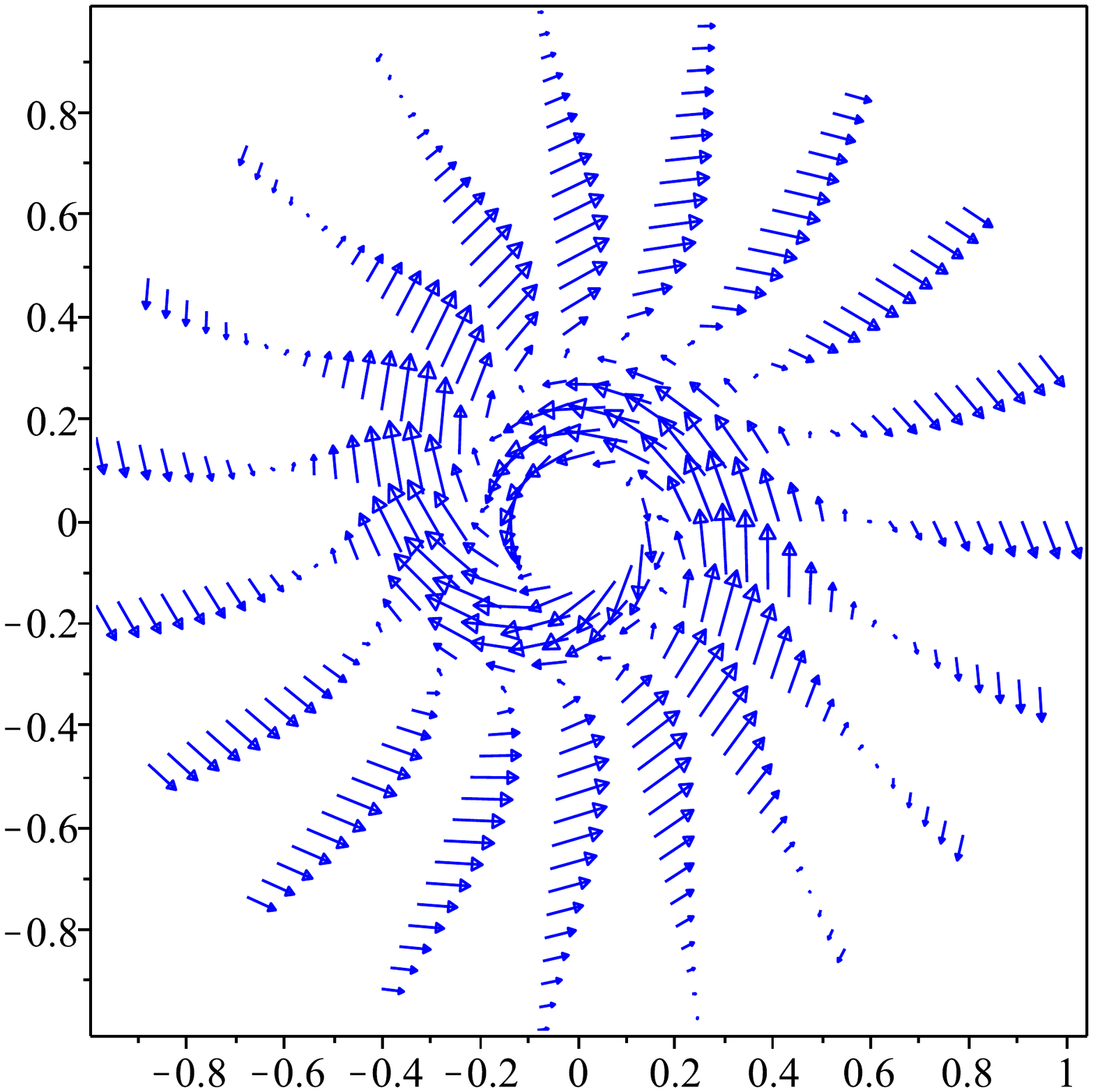}}}&
\rotatebox{0}{\scalebox{0.4} %change the angle and scale as you need
{\includegraphics{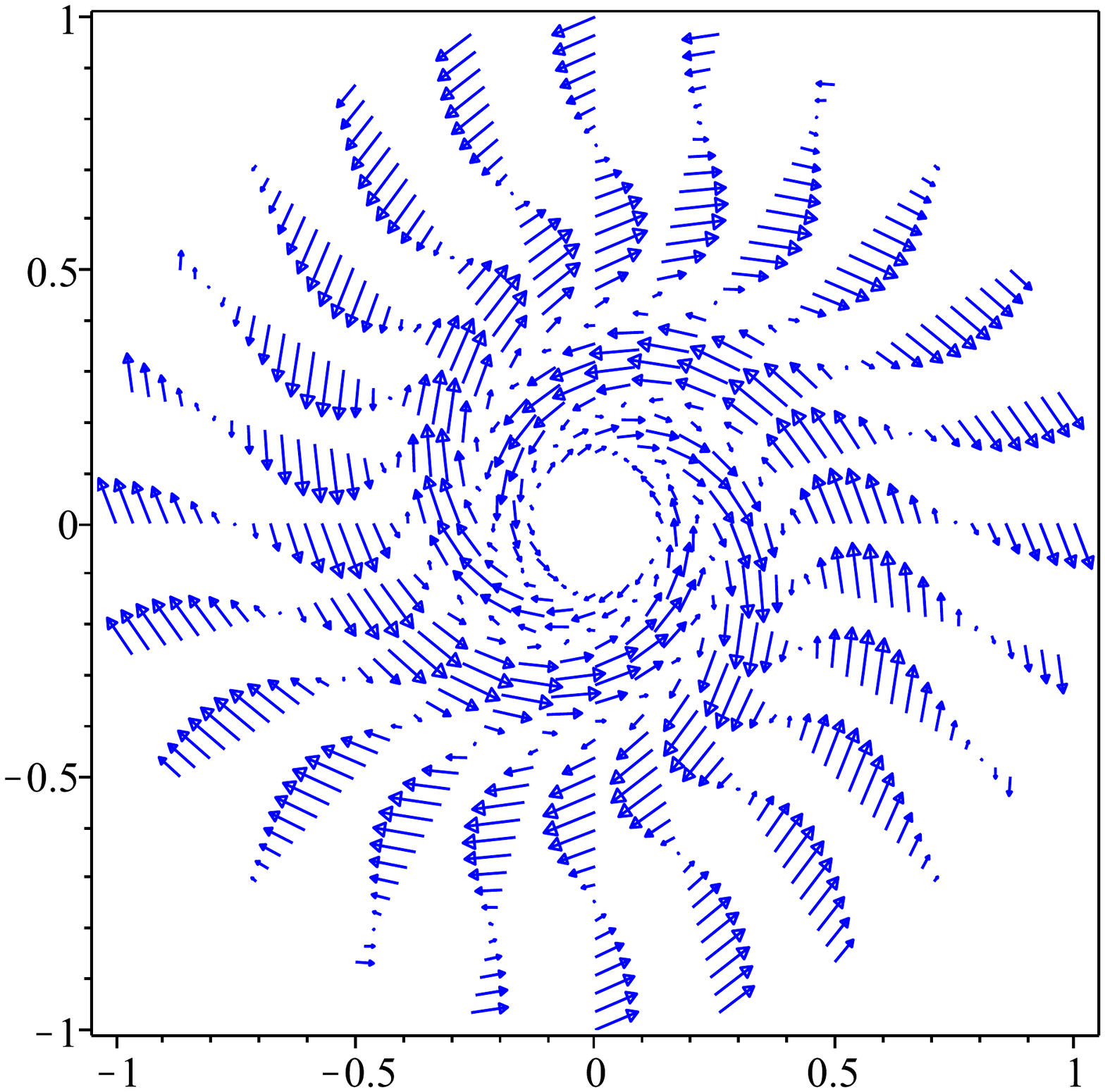}}}\\
{\rotatebox{0}{\scalebox{0.4} %change the angle and scale as you need
{\includegraphics{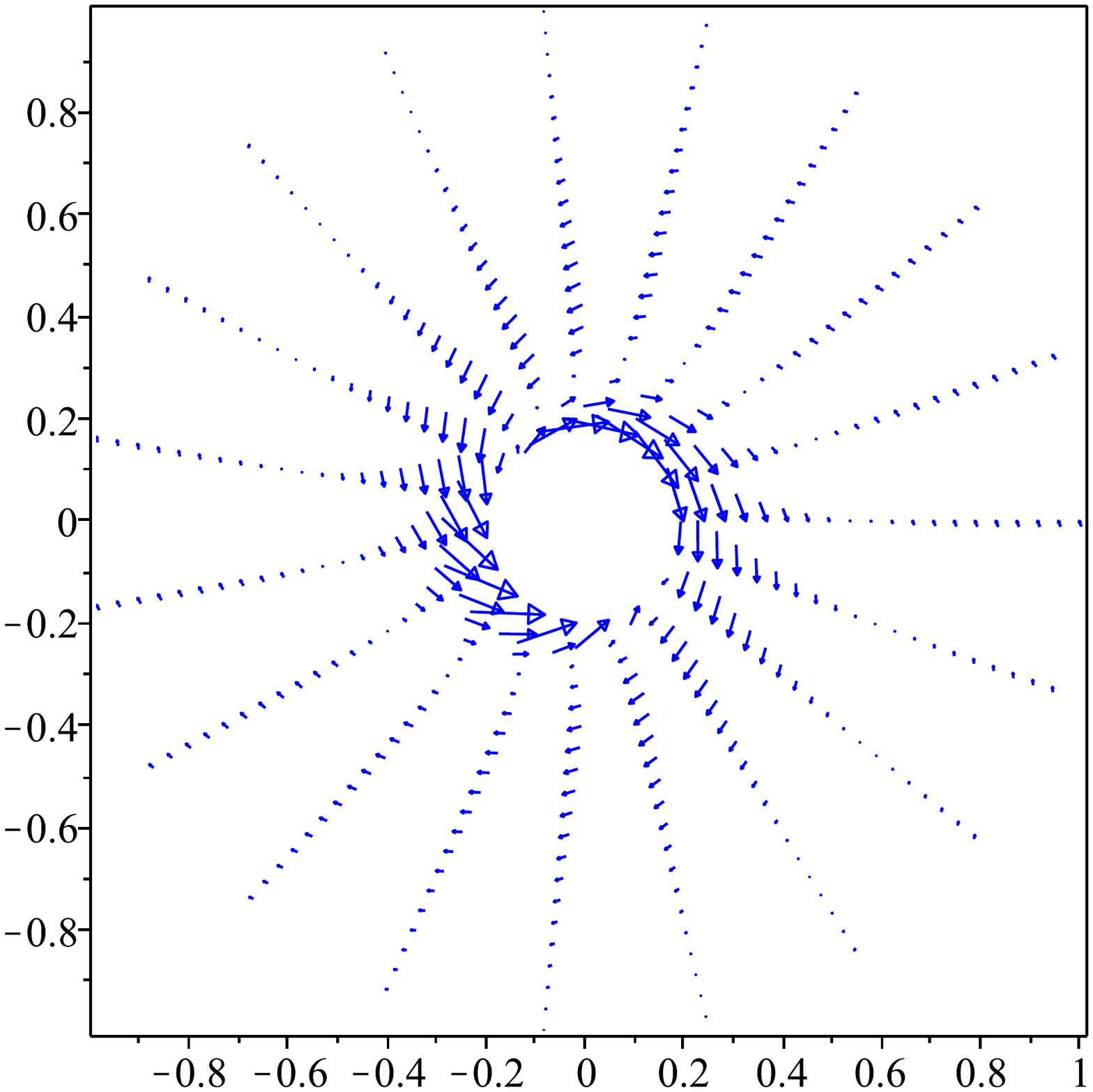}}}}&
\rotatebox{0}{\scalebox{0.4} %change the angle and scale as you need
{\includegraphics{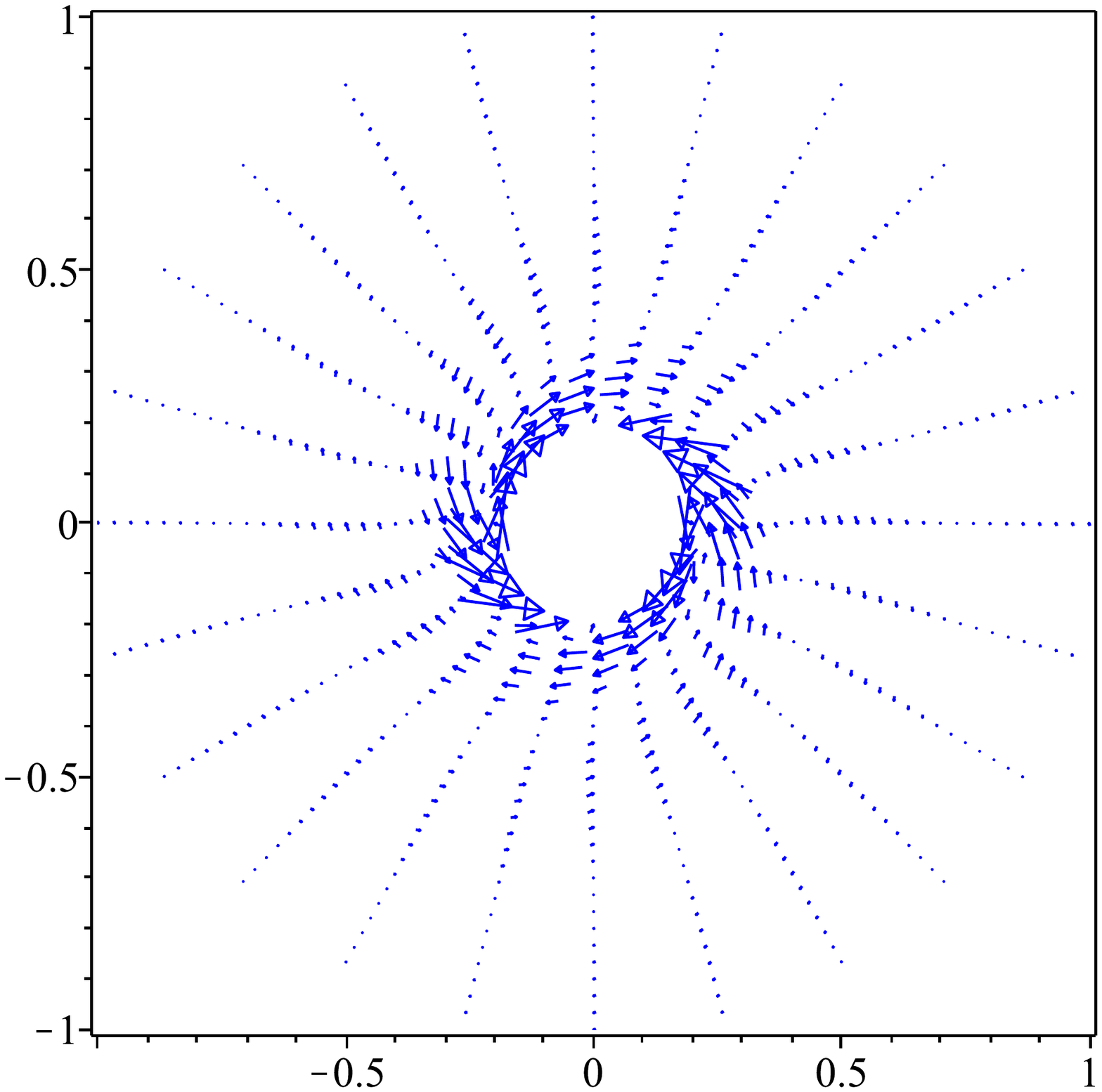}}}
\end{tabular}
\caption{At upper left we show the magnetic spiral that results from a cut through the halo magnetic field at $z=+0.1$, when $m=-1$, $q=0.4$ with $r=0.14..1$ and $\phi=0..2\pi$. We have set $s1=-1$ so that this case has an inflow with negative helicity.  At upper right the inflow  spiral structure for $m=-2$ at the same height, for the same parameters ($s1=-1$) is shown.  The helicity is also negative. At lower left the image is of the $m=+1$ arm with $s1=-1$ so that there is outflow and positive helicity. The height is the same as above but the vectors are now averaged, with the maximum vector reduces by $0.25$. The radius runs over $\{0.2,1\}$. At lower right we show the cut through the halo magnetic field at $z=0.1$  when there is also outflow. This shows the spiral from the $m=+2$ mode with $s1=-1$ , having the same parameters and the same presentation as the figure on the left.  The vectors in both lower figures are anchored on the head.    }    
\label{fig:Wonlydynamo}
\end{figure}

 Figure (\ref{fig:Wonlydynamo}) illustrates the behaviour of the magnetic field when only a vertical velocity field is present (recall that it falls off as $1/r$). The magnetic spiral, resulting from a cut through the halo magnetic field at $z=0.1R$, shown at upper left is strikingly well defined as a `magnetic arm' . The mode is  $m=-1$ and $s1=-1$. so that $\Delta<0$ and hence  the helicity is also negative. The gas is flowing into the galactic plane in this example. The contrast with the polarization arm from the same cut  seen at upper left in figure (\ref{fig:vzerodynamo}) is marked. True magnetic arms are never  globally produced in the zero velocity case, whatever the sign of the helicity. {\it This appears to be an important physical distinction favouring an $\alpha-\omega$ dynamo}.

\begin{figure}%[ht]
\begin{tabular}{cc} %This will make a two-column figure
\rotatebox{0}{\scalebox{0.6} %change the angle and scale as you need
{\includegraphics{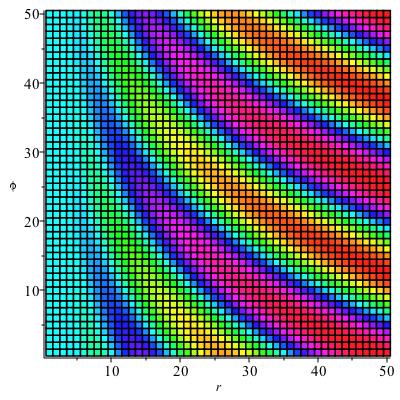}}}&
\rotatebox{0}{\scalebox{0.6} %change the angle and scale as you need
{\includegraphics{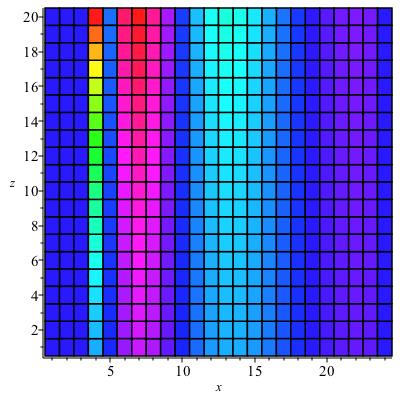}}}\\
{\rotatebox{0}{\scalebox{0.6} %change the angle and scale as you need
{\includegraphics{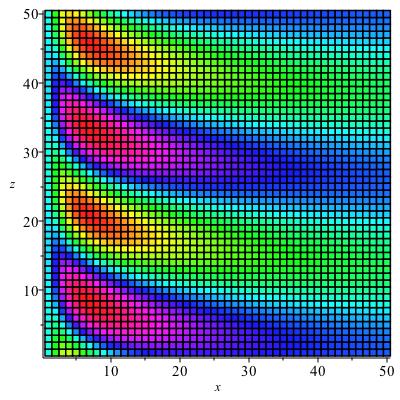}}}}&
\rotatebox{0}{\scalebox{0.6} %change the angle and scale as you need
{\includegraphics{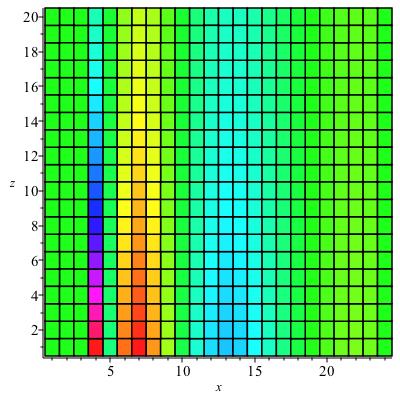}}}
\end{tabular}
\caption{The face-on RM assuming only an $m=-2$ accretion mode to be present ($s1=-1$. Other parameters are as in earlier presentations of this mode. The abscissa runs over $0.02\le r\le 1$, but the ordinate runs in azimuth from $0.125$ radians to $2\pi$ radians. Thus the top is continuous with the bottom. Each vertical half corresponds to one side of a face-on galaxy. The arms begin at small radius and large azimuth and trail to small azimuth and large radius. The bottom left figure shows the one armed accretion mode with $m=-1$ and $s1=-1$. The right two figures show the first and fourth quadrants for the outflow mode  from an {\it edge-on} galaxy with $m=-2$, and $s1=\pm 1$ according as $z\pm 0$. The ordinate  runs over $\{\pm 0.02,\pm 0.4\}$, while the abscissa runs over $\{1,6\}$. The radius of the disc is $6$ physical Units. The grid is blanked for $x<4$ grid Units. In grid Units for $z>0$ we have $(4,20)=-184.5,~(7,20)=+87.73,~(23,4)=11.24,~(14,8)=-38.5$. The fourth quadrant is antisymmetric.}    
\label{fig:faceonWRM}
\end{figure}

  On the right of the figure we show the two-armed spiral ($m=-2$, $s1=-1$)  resulting from the same cut and having  the same parameters.  The  accretion  and the negative  helicity ($\Delta=-ms1$ )  has produced  again very well defined magnetic arms . The cut is at $z=0.1R$, where $R$ is the disc radius. In the two bottom figures of figure (\ref{fig:Wonlydynamo}) we find one and two-armed spiral cuts with outflow ($m>0$  but $s1<0$ for $z>0$). 
  In these cases the magnetic field is increasing with increasing $z/r$, perhaps being carried up by the outflowing wind. The helicity is positive.  They are true magnetic spiral arms. The major difference from  accretion fuelled arms, is that these arms vary extremely rapidly in strength with radius.  Rotation measure variations would then be very difficult to detect as we illustrate ultimately  in figure (\ref{fig:faceonWRM}) for the edge-on outflow.

\begin{figure}%[p]
\begin{tabular}{cc} %This will make a two-column figure
\rotatebox{0}{\scalebox{0.6} %change the angle and scale as you need
{\includegraphics{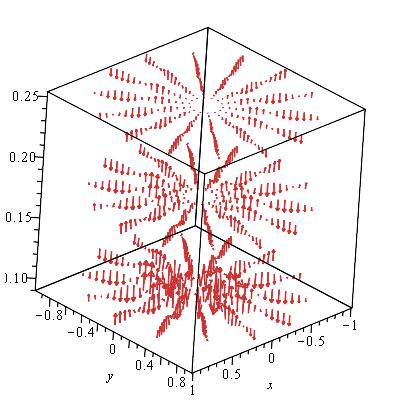}}}
\rotatebox{0}{\scalebox{0.6} %change the angle and scale as you need
{\includegraphics{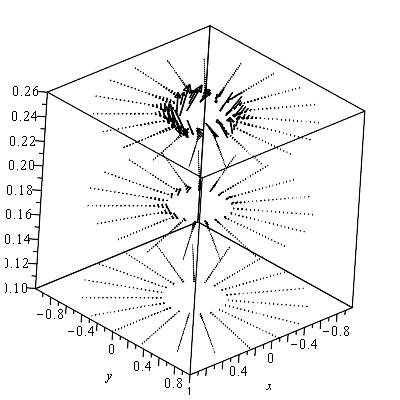}}}
%{\rotatebox{0}{\scalebox{0.8} %change the angle and scale as you need
%{\includegraphics{dynamosteadyfieldplot3dSp-1+.eps}}}}&
%\rotatebox{0}{\scalebox{0.8} %change the angle and scale as you need
%{\includegraphics{dynamosteadyspacecurveSp180-1+.eps}}}
\end{tabular}
\caption{At  left we show the field vectors in three dimensions  for $m=+2$, $s1=+1$ and $q=0.4$. This is the two-armed spiral case with inflow and negative helicity.  At  right  we show the field vectors for the outflow case with positive helicity having $m=2$ and $s1=-1$. The parameters are the same as on the left as is the presentation. We have allowed $0.15\le r\le 1$ and $ 0.1\le z\le 0.25$. }    
\label{fig:fieldplot3dW}
\end{figure}

\begin{figure}%[p]
\begin{tabular}{cc} %This will make a two-column figure
\rotatebox{0}{\scalebox{0.6} %change the angle and scale as you need
{\includegraphics{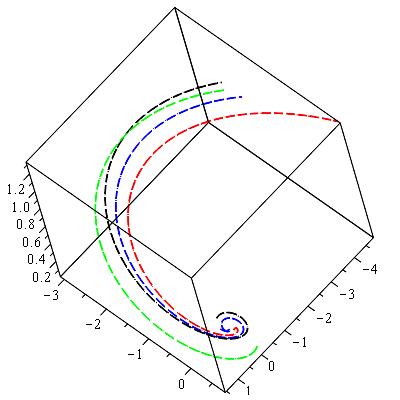}}}&
\rotatebox{0}{\scalebox{0.6} %change the angle and scale as you need
{\includegraphics{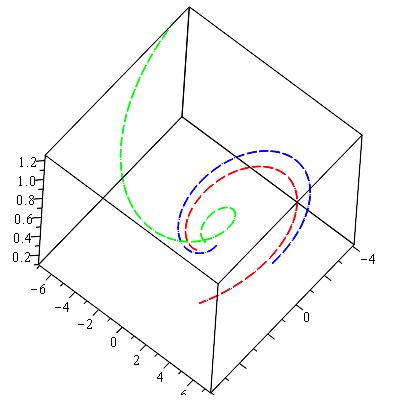}}}\\
{\rotatebox{0}{\scalebox{0.6} %change the angle and scale as you need
{\includegraphics{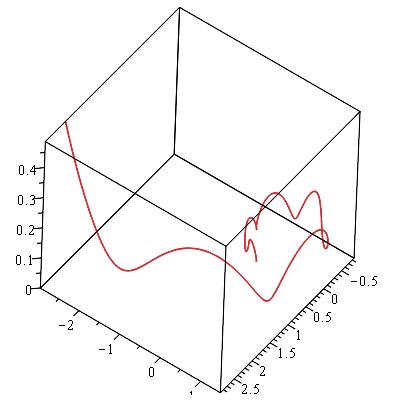}}}}&
\rotatebox{0}{\scalebox{0.6} %change the angle and scale as you need
{\includegraphics{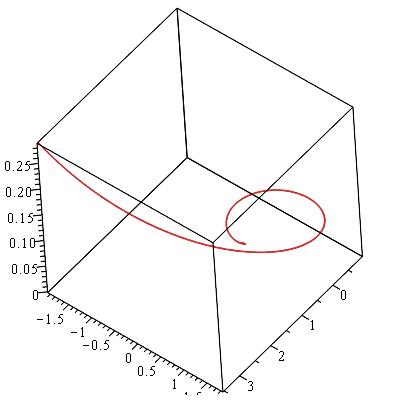}}}
\end{tabular}
\caption{ The upper left image shows a cluster of magnetic field lines for the one-armed accretion case with $m=-1$, $s1=-1$. The pitch angle is $\arctan{(0.4)}$. All of the field lines are picked up at $z=0.1$ and three of them begin on the radius $r=0.2$ at the angles $0$, $\pi/2$ and $3\pi/2$. The fourth line begins at a radius of $r=0.75$ with the azimuth $\phi=\pi/2$. The image at upper right shows three field lines for the two-armed accretion case with $m=-2$ and $s1=-1$. The pitch angle is the same as on the left image. Two of the lines begin at the radius $r=0.75$ at angles $0$ and $\pi/2$. The third line begins at $r=0.2$ and azimuth $0$. All lines start at $z=0.1$.  The field line at lower left is for the two-armed spiral with accretion, but $m=+2$ and $s1=+1$ at $z>0$. It begins at $z=0.0001$, $r=0.2$  and $\phi=0$. The field line at lower right is the two-armed spiral with accretion and $m=-2$ and $s1=-1$. Nothing else is changed relative to the image at lower left.}
\label{fig:spacecurveW}
\end{figure}

The difference between  two-armed halo magnetic fields with inflow or outflow is explored in three dimensions in figure (\ref{fig:fieldplot3dW}). On the left we have the field vectors for the mode $m=2$ at $z=0.1$ to $0.25$. The pitch angle is  $22^\circ$ as usual and there is accretion with negative helicity. The field is complicated in the extreme, with what appear to be looping field lines over the arms seen in the cut shown at upper right in figure (\ref{fig:Wonlydynamo}). We show this  complicated behaviour for one line in figure (\ref{fig:spacecurveW}) near the disc.  

The contrast with the outflow (and positive helicity) case on the right of figure (\ref{fig:fieldplot3dW}) is of considerable  interest. The outflow field is simpler but grows rapidly with height at small radius.  This is the source of excessive contrast in the rotation measure. This is in general agreement with the idea that the outflow removes magnetic flux from the galactic disc. Although it may not be evident at this angle in the figure, there is an `X type' projected magnetic field in this image. It is likely to be the $m=0$ field that dominates `X type' behaviour however.

Figure (\ref{fig:spacecurveW}) integrates some magnetic field lines using the same  field line equations as in the zero velocity example, but with the magnetic fields adapted to this case. At  upper left we show a cluster of field lines for the mode $m=-1$, $s1=-1$ inflow. All of the lines begin at a height of $z=0.1$ (in Units of the disc radius) . Three of the lines begin at $r=0.2$ at angles $0$,$\pi/2$ and $3\pi/2$. The fourth line begins at $\phi=\pi/2$ and $r=0.75$. Even though the full height reached by the lines exceeds our approximation limit, the indication is of a smooth spiralling magnetic field rising from the disc and wrapped on cones.

At upper right the cluster of field lines is for the two-armed accretion $m=-2$ and $s1=-1$. They are identified passing through the height $z=0.1$. Two begin at radii $r=0.75$ at angles $0$ and $\pi/2$, and the third begins at $r=0.2$ at zero angle. Once again the indication is of a magnetic field rising smoothly on conical helices. This holds even if one begins the field line very close to the disc, as is demonstrated  for this mode at lower right.  This is in remarkable contrast to the field line at lower left, which is found in the $m=+2$, $s1=+1$  ($z>0$ ) inflow mode. The line begins at $r=0.2$, $\phi=0$  and $z=0.0001$, the same height as for the image at lower right.  Here the change in mode sign has produced a magnetic field close to the disc that loops over the disc extensively, before moving off to greater heights. The loops are sufficiently near to the disc that our approximation should  hold. Thus loops of this type would give small scale oscillating rotation measures, which can imitate Parker instability.  It may actually physically imitate the Parker instability because we do not detect this behaviour without inflow.

The field lines for the outflow modes are rather complicated and need an expanded study to do them justice. Low loops starting near the galactic plane seem common, however.

\begin{figure}%[p]
\begin{tabular}{cc} %This will make a two-column figure
\rotatebox{0}{\scalebox{0.6} %change the angle and scale as you need
{\includegraphics{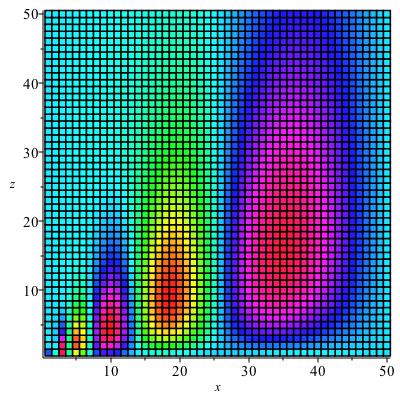}}}&\\
\rotatebox{0}{\scalebox{0.6} %change the angle and scale as you need
{\includegraphics{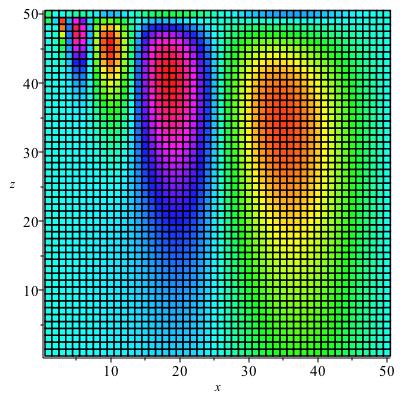}}}&
%{\rotatebox{0}{\scalebox{0.8} %change the angle and scale as you need
%{\includegraphics{dynamosteadyfieldplot3dSp-1+.eps}}}}&
%\rotatebox{0}{\scalebox{0.8} %change the angle and scale as you need
%{\includegraphics{dynamosteadyspacecurveSp180-1+.eps}}}
\end{tabular}
\caption{We show the rotation measure in the first and fourth quadrants for the infall two-armed mode $m=+2$, $s1=+1$. The second and third quadrants can be found by interchanging colours after rotation according to the left-hand rule about the left edge (the galactic axis). The sign significance of the colours should also be interchanged. The abscissa runs over $0.02\le r \le 1$ while the ordinate runs over $0.01\le z \le 0.5$.  }    
\label{fig:RMplotW+2+1}
\end{figure}

We look once again at the `rotation measure' (RM)  produced in the halo by this dynamo magnetic field. As in case A we really  calculate the isolated effect of the  dynamo magnetic field  in a Faraday screen with constant electron density. Figure (\ref{fig:RMplotW+2+1}) shows the two-armed halo with $m=2$, $s1=+1$ infall. As before the halo is cylindrical with the disc radius, and the inclination of the galaxy is $90^\circ$. The second and third quadrants maybe found by rotating about the left edge of the figure according to the left-hand rule. Subsequently the colours are interchanged. The electron density is constant, but it should be used to cut off the behaviour in $z$ , which rises to the limits of our approximation. The horizontal axis runs from   $r=0.02$ to $r=1$ as before, while the vertical axis runs from  $z=0.01$ to $z=0.5$.  The antisymmetry is evident on crossing the plane.The Unit is one disc radius.

\begin{figure}%[p]
\begin{tabular}{cc} %This will make a two-column figure
\rotatebox{0}{\scalebox{0.6} %change the angle and scale as you need
{\includegraphics{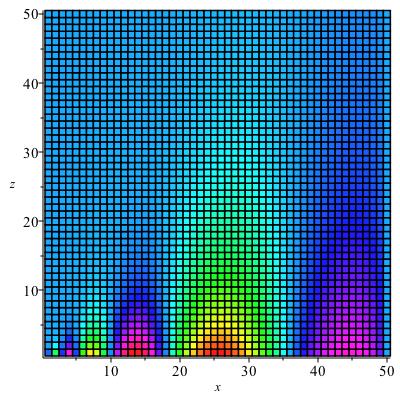}}}&\\
\rotatebox{0}{\scalebox{0.6} %change the angle and scale as you need
{\includegraphics{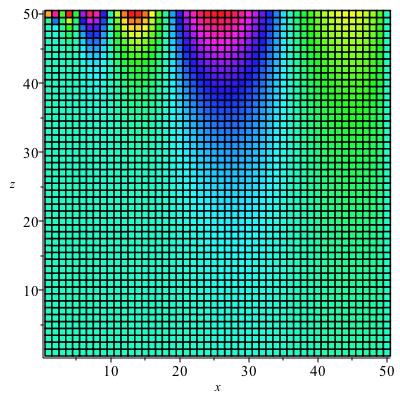}}}&
%{\rotatebox{0}{\scalebox{0.8} %change the angle and scale as you need
%{\includegraphics{dynamosteadyfieldplot3dSp-1+.eps}}}}&
%\rotatebox{0}{\scalebox{0.8} %change the angle and scale as you need
%{\includegraphics{dynamosteadyspacecurveSp180-1+.eps}}}
\end{tabular}
\caption{We show the rotation measure in the first and fourth quadrants for the inflow two-armed mode $m=-2$, $s1=-1$. The second and third quadrants can be found by interchanging colours after rotating according to the left-hand rule about the left edge (the galactic axis). The sign significance of the colours should also be interchanged.  The abscissa runs over $0.02\le r \le 1$ while the ordinate runs over $0.01\le z \le 0.5$.  }    
\label{fig:RMplotW-2-1}
\end{figure}

In figure (\ref{fig:RMplotW-2-1}) we show the  rotation measure for the  two-armed inflow mode $m=-2$, $s1=-1$.  The peak RM is much closer to the plane compared to the positive mode in figure (\ref{fig:RMplotW+2+1}). The same interchange symmetry and rotation allows the construction of the second and third quadrants as previously. 

Compared to the mode $m=+2$, $s1=+1$ mode found in case A, figure (\ref{fig:RMplotW+2+1}) shows smaller azimuthal magnetic field near the disc.
This is consistent with the low altitude looping field seen in figure (\ref{fig:vzerodynamo}) and with the smooth, slowly spiralling field lines seen in figure (\ref{fig:spacecurveW}).

\newpage
Finally  having the halo magnetic fields consistent with a steady dynamo  with outflow allows us to consider the `rotation measure' (i.e. Faraday screen) as seen for a face-on galaxy. We assume it to be perfectly face-on and integrate $b_z(r,\phi,z)$ over $z$ from $-0.25$ to $+0.25$, taking into account the symmetry of the vertical field above and below the plane. The calculation essentially sees a halo Faraday screen produced by the halo in front of the disc emission. We show a preliminary plot in figure (\ref{fig:faceonWRM}) of the inflow mode $m=-2$, $s1=-1$ . 
The abscissa is again the radius running from $r=0.02$ to $r=1$. The vertical axis however is the azimuthal angle running from $0.1256$ radians to $2\pi$ radians. The image is then approximately topologically periodic with top edge to be joined to the bottom edge. This can be done by rolling the figure about the horizontal  mid-line. Each half of the figure in the vertical sense corresponds to half the galaxy on the sky.

The spiral magnetic structures of alternating sign run from large   azimuth at small radius to large radius and smaller azimuth. However the location of the zero azimuth on the galaxy is an arbitrary choice. The cause must be attributed to the looping magnetic field associated with the spiral arms. One expects then an anti correlation with the polarized intensity, and indeed with the location of the magnetic arms  as seen in face-on galaxies.  Distinct  rotation measure patterns such as these are likely to be confused by local stellar sources of  magnetic field and /or electron density. We can hope that global averages may be less sensitive to  such perturbations.  The mode $m=-1$, $s1=-1$ is similar but has only one rotation measure arm (of opposite sign) on each side of the minor galactic axis. This behaviour is in addition to the small scale RM oscillations to be expected for infall modes.

Remarkably such behaviour may have already been observed in IC342 \cite{Beck2015}. On the left of figure $26$ of that paper we find the RM plotted over one half of the face-on galaxy. The  observed oscillation is much as we expect qualitatively  from  the left images in our figure (\ref{fig:faceonWRM}) . The observation will have to be fit quantitatively, but the fact that the positive and negative amplitudes are observed to be similar is typical of the model result. Moreover on the right of figure $26$ we find the polarized intensity (PI) plotted over the same sector. At least where the RM is largest, we see the expected anti correlation with the PI. This must be regarded as encouraging. Figure $27$ of that paper shows a striking  $m=2$ global magnetic spiral  
much as we find in our outflow/inflow model.

The right-hand side of figure (\ref{fig:faceonWRM}) shows the outflow mode $m=-2$ and $s1=\pm 1$ (depending on whether $z=\pm 0$) from an {\it edge-on} galaxy. The contrast with the previous inflow modes is marked. The radius (abscissa) runs over $1\le r\le 6$ where the disc radius is $6$ Units. We have taken the upper limit in $|z|$ to be $0.4$, which is also the maximum in $Z$. The second and third quadrants can be found by rotation about the left edge of the figure according to the left-hand rule and subsequently interchanging the colours. The grid is only $20\times 20$ in this image because the first four units on the abscissa should be blanked.

{\it The important difference from the accretion modes  shown earlier is the strong increase of magnetic field with height associated with the outflow.} The figure should be compared to the inflow RM found in figure (\ref{fig:RMplotW-2-1}). Had we continued this figure to a tenth of the galactic radius rather than one sixth as in the figure, one would find very strong RM at large $z$. This would render the rest of the oscillations difficult to detect. Hence we must look for the behaviour shown in the figure at large radius so that $Z$ is small. The current contrast is already seen to be large between the upper left  and lower right of the figure for positive $z$.

\newpage

\section{Discussion and Conclusions}

 Using scale invariance symmetry we have expanded steady mean field dynamo solutions in logarithmic spiral modes. The results may be considered to hold either in an inertial frame, or in a locally rotating `pattern' frame. The application of these solutions above the galactic disc is restricted to regions  where the tangent of the (complementary) conical angle $z/r$ is small, but an exact solution agrees in the over lap region.
 
 The scale invariance also substitutes for the dynamics in this model. However one should recall that systems far away from initial conditions and spatial boundaries are often in a scale free form. Thus, although the dynamo fields are kinematic, the power law velocities may not be so unrealistic.  Moreover the helicity and diffusivity spatial dependence is also dictated by the scale invariance. {\it Our results are independent of arbitrary dependences on $\kappa$ of the helicity and and diffusivity, provided these are the same. This dependence is also communicated to the physical velocity through the diffusivity.} It is not unreasonable that these dependences should be the same if they are due to sub-scale turbulence.

%This work is complemented by a separate study of the axisymmetric mode in paper I (Henriksen, 2017). A simplified study  (no $\alpha$ effect) was done for a diffusing and shearing magnetic field in \citet{HI2016},  provided that the velocity is parallel to the magnetic field. 
%The necessary force equilibrium to maintain the steady state was found there by assuming dynamical equipartition. 

%This study  avoids the actual dynamics that must be associated with the dynamo action, so that it is a kinematic study only wherein the velocity is given in form by scale invariance and the amplitude is a parameter. 

The gravitational field  is an essential part of the ultimate dynamics and  in Appendix B we have given the exact gravitational field of a logarithmic spiral in a uniformly rotating pattern reference frame following \citet{Kal1971}. To  these spiral perturbations one must ultimately add an axisymmetric  gravitational mode, perhaps approximated by an isothermal dark halo and a Mestel disc as in \citet{HI2016}. The addition of magnetic, inertial and pressure gradient forces will complicate the dynamical problem. It seems that this system has not yet been simulated elsewhere.

 One should note that the new perspective  presented in this Appendix is that this gravitational spiral is also scale free (cf \cite{BT2008}). This perspective  allows some speculation regarding the relative evolution of the gravitational and magnetic spiral arms.  In particular there is a prediction that the magnetic and gravitational arms may cross in some cases.

Our approximate formulation for the scale invariant magnetic dynamo is summarized in equations (\ref{eq:dynamor}), (\ref{eq:dynamophi}) and (\ref{eq:dynamoz}). These require $Z=z/r$ to be $<1$  for validity, as  is shown in Appendix A, but they allow general parameters.  They are a function of the single variable  $\kappa$ (\ref{eq:trans1}).  A more general scale invariant form requires $Z$  and $\xi$ to be  independent variables and leads to the partial differential equations given in appendix A.  

For a particular choice of radial velocity and `dynamo' number, an exact solution was found. This solution justifies our approximate solution in the over lap region and extends the spiral magnetic fields well into the halo. In general this can not be done.

 Under the approximation the modal and scale invariant ans\"atze lead directly to the homogeneous modal equations. The vanishing determinant of these equations yields our major constraint on a steady mean field dynamo in equation   (\ref{eq:detA}). This gives the Reynolds number of the sub-scale turbulence  (i.e. the dynamo number) that is required for the existence of a steady dynamo as a function  of the {\it scaled} velocities (also parameters), the magnetic spiral pitch angle, and the mode number. Subsequently we studied  three particularly simple special cases for their observational consequences.  
 
 The steady dynamo with zero velocity in the pattern frame or with  velocity parallel to the magnetic field was our first example. 
 
  % We emphasize that  that {\it no dynamics} is required to solve the classic dynamo in this case. One is simply placed in a pattern frame where there is  no peculiar velocity. 
  
  The Reynolds (dynamo) number is equal to the mode number multiplied by $s1$.  A negative  value implies that the sub-scale helicity is negative.  We have looked at cases with both positive and negative helicity. In neither case are true global  magnetic spiral arms produced, rather they are polarization arms. This is illustrated in figures  (\ref{fig:vzerodynamo})  and (\ref{fig:vzeroField3d}).
 
 Figure (\ref{fig:vzeroField3d}) also shows the magnetic vectors near the plane ($Z$ small) in three dimensions. There is a strong indication of magnetic loops associated with the magnetic spiral arms for the $m=2$ mode.  Two such loops are illustrated in figure (\ref{fig:vzerodynamo}). These are characteristic of the steady alpha-squared  turbulent dynamo. These loops are seen at small radius  to close  near  the plane.  Nevertheless they extend over large regions as a mean field, and  they might be detected with suitable data smoothing at larger radii.  
 
 % Oscillating Faraday rotation measures in face-on galaxies can be expected based on these loops. Oscillations  similar to those produced in edge-on galaxies by spiral field lines oscillating in azimuth should be produced. They should show the same anti-correlations with polarized intensity when present. This scenario  seems most likely in the presence of inflow.
 The bottom images in figure (\ref{fig:vzeroField3d}) are of a $m=2$, $s1=1$  accretion mode on the left and $m=-2$, $s1=-1$ accretion on the right. We see mainly global polarization arms in each case.
 
 Each  spiral mode $m$ must ultimately be combined with a related axially symmetric ($m=0$) solution. In the combination we expect both the `X-type' fields seen in polarization, plus oscillations in rotation measure  (RM) and polarization intensity, to be produced. Halo lag  as in (\cite{HI2016}) may also be produced in this combined field, on using the appropriate boundary conditions at the disc and at infinity.  
 
{\it It is characteristic of the magnetic fields produced by the steady mean field dynamo of this paper, that the tangential fields change sign on crossing the plane} (\ref{fig:naiveRM}), (\ref{fig:RMplotW-2-1}).  This allows the perpendicular field to be continuous there.  In the axially symmetric field, it is possible to have a kind of `quadrupole symmetry' wherein $b_z(r,\phi,0)=0$.  This is not possible for these non-axially-symmetric modes. 
 
 In order to illustrate what might be possible observationally, we have calculated in figure (\ref{fig:naiveRM}) the `rotation measure' (Faraday screen)  given by our halo magnetic fields when integrated along the line of sight. The electron density is held constant in order to isolate the mean field magnetic effect.  
  Nevertheless the figure illustrates a new  observational possibility, that of oscillating RM sign {\it on the same side of  the galactic minor axis}.  Only the contrast  in amplitude between the positive and negative RM peaks is fixed by our calculation, because there is an arbitrary multiplicative constant in the magnetic field.  The angle by which the RM peaks emerge from the plane indicates substantial azimuthal field at the disc, originating in the closed loops of figure (\ref{fig:vzerodynamo}).  
  
We note that this lifting of spiral arms into the halo of the galaxy has  also been found in earlier work \cite{MBDT93}, although the observational consequences were not  greatly elaborated. Figures $4$, $5$ , $9$ and $10$ of that paper are relevant to the present work. We note that they used the {\it same} classical dynamo theory as do we, except that they follow its  temporal evolution, beginning from some arbitrary `seed' magnetic field. There are certainly times where the magnetic field resembles that found in our steady state approach. Figure $5$ indicated that field anti-symmetry across the equator that we also find. Figures $4$ and $9$ illustrate the transitory lifting spirals.  Figure $10$ illustrates the sign changing RM {\it in the equatorial plane.} The axially symmetric field seems to dominate asymptotically, which is the subject of our companion paper\cite{Hen2017}. These authors obtain only a qualitative fit to the magnetic field of M81, as is our current objective.

Very similar results are found in the paper \cite{DB90}. Once again rather particular assumptions are made about helicity and diffusivity and angular velocity. The field behaviour (particularly the field lines) is very similar to what we find here. These authors have in addition calculated the RM ,much as we have done, {\it but lying along the galactic plane}. The behaviour predicted is compatible with our own predictions, which in our case extend above the plane.

All these authors have used, physically motivated but arbitrary, forms of the variation in diffusivity, helicity and halo velocity. These  quantities, coupled with the uncertain seed field, suggest that  asymptotically these parameters are not so important.  We have avoided many of them be assuming a steady state and scale invariance as the asymptotic state.

 Our conclusions based on this example are:
 
 $\bullet$  Magnetic Spiral Arms produced by the pure alpha-squared dynamo are mainly polarization arms.
 
 $\bullet $ The magnetic field structure associated with the galactic plane continues into the halo. Disc spiral structure is `lifted' onto cones.
 
 $\bullet $The spiral structure is naturally anti-symmetric across the galactic disc.
 
$ \bullet$  Observationally, oscillations are expected in the RM   measures of edge-on galaxies, even on the same side of the galactic minor axis.  Face-on galaxies should exhibit  in the mean field, spiral structure in the RM. These variations may well be associated with oscillations in the polarization intensity.

 A second special case was that of a simple $\alpha-\omega$ dynamo when only an azimuthal velocity in a pattern frame was present, Unlike the zero velocity case, true magnetic spiral arms are readily produced with magnetic field decreasing with height above the disc ($ms1>0$). Increasing magnetic field ($ms1<0$) gives either very strongly contrasted magnetic spirals ($m=1$) or strong polarization arms ($m=2$). The magnetic field near the disc consists of high loops, which, in the increasing field case pass through the centre of the galaxy (cf \cite{DB90}). In this way they behave similarly to the field of a point magnetic dipole.  The RM oscillation predicted for an edge-on galaxy is much as that found for the inflow case discussed below. 
 
 This case can be used to combine outflow with rotation, and we gave the necessary parameters in a corollary. However we leave the details of this dynamo to future work.  
 
 The third  special case  is when only a z velocity component $w$ is present. The sub scale Reynolds number is equal to minus the mode number times $s1$, and the mode number is in turn given by $ms1=-q w/2$. Thus vertical outflow ($w>0$) requires $ms1<0$ above the plane and $ms1>0$ below the plane.  Figure (\ref{fig:Wonlydynamo})  shows at upper left a well defined magnetic spiral for $m=-1$ with inflow to the disc.  At upper right this is repeated for two-armed inflow. An $m=+1$ spiral  with outflow ($s1=-1$) (lower left image) also yields  a well defined magnetic arm . However it is much more concentrated at small radius and large (within the bounds of our approximation) $z$.  This is confirmed in the image at lower right for a two-armed outflow ($m=2$,$s1=-1$).

   We see in all of these images well-defined true magnetic arms, which would project to the magnetic arms that have been detected in face-on galaxies. The outflow modes are less radially extended than the accretion modes  but grow in $z$. They have positive helicity. The accretion modes are relatively compressed and have negative helicity.   
 
 Figure (\ref{fig:fieldplot3dW}) shows the inflow and outflow vectors ($m=\pm 2$, $s1=-1$) . We see that the looping is evident in the inflow case, but that the outflow is much smoother at theplane.  Moreover, the magnetic field grows markedly in the outflow case. 
 
  The field line plots  in figure (\ref{fig:spacecurveW}) are for the more accessible accretion modes. The change in sign of the mode number has an interesting effect on the field line loops. 
 We see that the spiral  accretion fields mostly rise smoothly on helical cones into the galactic halo. The accreting two armed flow with positive mode number at lower left produces a disturbed looping field near the disc, which only eventually escapes to infinity.

The figures showing the RM  (\ref{fig:RMplotW+2+1}) indicate some differences with those of the pure $\alpha^2$ example.  The inflow pattern shows the RM peaks lifted away from the disc in a roughly conical distribution. This appears to be due to the delivery of magnetic flux from beyond the disc (see figure (\ref{fig:Wonlydynamo}) for the enhanced compactness of the inflow spiral cuts). The negative mode  spiral pattern (\ref{fig:RMplotW-2-1})
is more similar to that of the previous section. However it lacks the strong azimuthal field component that the looping field provides for the alpha-squared dynamo. The RM pattern is confined to the near disc region by the inflow.

Similar effects may be expected for face-on galaxies. 
Figure (\ref{fig:faceonWRM}) represents the spiral RM pattern expected for a face-on galaxy. All of these RM calculations must be made more realistic, particularly as to the expected amplitude, but the general fact of oscillation  is  due to the closed loops that leave the plane.  One should remember that these magnetic fields are produced by a mean-field dynamo. As such only suitable averages can be expected to behave in this manner. We show finally in this figure the  RM in the first and fourth quadrants for an outflow edge-on mode with $m=2$, $s1=-1$. The contrast with the accreting modes is marked, and shows an extension into the halo that may have been expected.

There seems to be a very clear observation of the effects anticipated in this work in the study of IC 342 in \cite{Beck2015}. Such effects have been suggested earlier (e.g. \cite{MBDT93}, \cite{DB90}), but mainly with the galactic plane in mind and including many parameters. Nevertheless the early work and the present study show remarkable convergence.

To our previous conclusions we can  on the basis of this example add the following:

$\bullet$ Outflow  certainly enhances the magnetic field structure, so long as $ms1<0$  (above the plane) and the magnetic field increases with $z/r$.  The helicity is positive. 

$\bullet$ Inflow produces more  compact magnetic structure, and requires negative helicity.  The magnetic spirals rise smoothly on cones into the halo under accretion ($ms1>0$ above the plane). 

$\bullet$ Inflow produces open magnetic field line loops near the plane. These lines eventually rise into the halo. Observationally, they may create small scale RM and polarized intensity oscillations. Strong looping magnetic field is produced by the pure $\alpha-\omega$ dynamo. The field line loops pass  
through the galactic centre when the field increases away from the plane.

$\bullet$ The RM distribution for edge-on galaxies with outflow extends farther into the halo than does the inflow distribution. Same side galactic minor axis  sign oscillations are expected in both inflow and outflow, but the outflow oscillations  are probably mainly visible at small $z/r$ because of excessive contrast. 

$\bullet$ The RM distribution for face-on galaxies should in the mean take the form of spirals alternating in sign. The polarization intensity should also oscillate in amplitude. These may be more difficult to observe because of the inevitable local structure and turbulence (necessary for the dynamo) found in the galactic disc. The amplitude may be weaker because of the shorter path length through the galactic disc. 
 
\section {Acknowledgements}

This work was undertaken as a visitor to the Astronomy Group at the R\"uhr Universit\"at Bochum,which is led by Prof Ralf-J\"urgen Dettmar who was a splendid host. The visit was financed by a research grant from the Alexander von Humboldt foundation. Reiner Beck gave many constructive comments on an earlier version of this draft. Marita Krause, Carolina Mora and Philip Schmidt are thanked for discussion. The author acknowledges a suggestion by Arpad  Miskolczi regarding polarization intensity variations.  Judith Irwin always gives wise advice.

\newpage
 \appendix
 \centerline{\bf Appendix A: Two Variable Self-Similarity}
 
 In this appendix the exact two-variable scale invariant equations are given and  are reduced to our approximate form. Both the vector potential and the magnetic field are used.

 We use the same variables as in the text. Equation (\ref{eq:Afield})  of the text then takes the three explicit forms under scale invariance: 
 
 The radial component;
 
 \bea 
 &\Delta&(q\partial_\xi\A_z-\partial_Z\A_\phi)+v((3-a)\A_\phi-Z\partial_Z\A_\phi+\partial_\xi\A_\phi-q\partial_\xi\A_r)\nonumber\\
 &-&w(\partial_Z\A_r-(2-a)\A_z+Z\partial_Z\A_z-\partial_\xi\A_z)-\big((3-a)q\partial_\xi\A_\phi-qZ\partial_\xi\partial_Z\A_\phi+q\partial_\xi^2\A_\phi\nonumber\\
 &-& q^2\partial_\xi^2\A_r-\partial_Z^2\A_r+(2-a)\partial_Z\A_z-\partial_Z(Z\partial_Z\A_z)+\partial_Z\partial_\xi\A_z\big)=0.\label{eq:AmultSSr}
 \eea
 
 The azimuthal component;
 
 \bea
& \Delta&(\partial_Z\A_r-(2-a)\A_z+Z\partial_Z\A_z-\partial_\xi\A_z)-u((3-a)\A_\phi-Z\partial_Z\A_\phi+\partial_\xi\A_\phi-q\partial_\xi\A_r)\nonumber\\
&+&w(q\partial_\xi\A_z-\partial_Z\A_\phi)+(3-a)\partial_\xi\A_\phi-(3-a)Z\partial_Z\A_\phi+Z\partial_Z(Z\partial_Z\A_\phi)-Z\partial_\xi\partial_Z\A_\phi\nonumber\\
&+&\partial_\xi^2\A_\phi-Z\partial_Z\partial_\xi\A_\phi-q\partial_\xi^2\A_r+qZ\partial_Z\partial_\xi\A_r-q\partial_Z\partial_\xi\A_z\nonumber\\
&+&\partial_Z^2\A_\phi+(1-a)((3-a)\A_\phi-Z\partial_Z\A_\phi+\partial_\xi\A_\phi-q\partial_\xi\A_r)=0. \label{eq:AmultiSSphi}
\eea

The vertical component;

\bea
&\Delta&((3-a)\A_\phi-Z\partial_Z\A_\phi+\partial_\xi\A_\phi-q\partial_\xi\A_r)+u(\partial_Z\A_r-(2-a)\A_z+Z\partial_Z\A_z-\partial_\xi\A_z)\nonumber\\
&-&v(q\partial_\xi\A_z-\partial_Z\A_\phi)-\big(\partial_Z\A_r-(2-a)\A_z+Z\partial_Z\A_z-\partial_\xi\A_z-q^2\partial_\xi^2\A_z+q\partial_\xi\partial_Z\A_\phi\nonumber\\
&+&(1-a)[\partial_Z\A_r-(2-a)\A_z+Z\partial_Z\A_z-\partial_\xi\A_z]-Z\partial_Z^2\A_r+\partial_\xi\partial_Z\A_r-(2-a)\partial_\xi\A_z\nonumber\\
&+&(2-a)Z\partial_Z\A_z-Z\partial_Z(Z\partial_Z\A_z)+Z\partial_\xi\partial_Z\A_z+Z\partial_Z\partial_\xi\A_z-\partial_\xi^2\A_z\big)=0.\label{eq:AmultiSSz}
\eea

These are the complete equations for what is two Dimensional Self-Similarity in terms of $\xi$ and $Z$. We have grouped the terms as they appear in the alpha dynamo, the global dynamo (velocity dependence) and the turbulent diffusion. 

At this stage we have  succeeded  only in removing the radial dependence by the assumption of scale invariance. Such scale invariance seems appropriate for the intermediate disc structure of galaxies. However the subsequent solution of this set of partial differential equations is a separate problem. One way of reducing the system to a set of three ordinary equations in $Z$ is to assume ${\bf A}={\bf C}\exp(ip\xi)$, and then to solve for ${\bf C}(Z)$ as a complex function of a real variable.  However this produces three  (six if split into real and imaginary parts) equations  for ${\bf C}$, each one of second order in $Z$, and there does not seem to be a simple solution. 

An apparently simpler approach to the exact problem is to change back from ${\bf A}$ to the scaled magnetic field components. These  scaled field components are 
\bea 
\bar b_r&=&\delta(-\partial_Z \A_\phi + q \partial_\xi \A_z),\label{eq:Abrbar}\\
\bar b_\phi&=& \delta(\partial_Z\A_r+(a-2)A_z+Z\partial_Z\A_z-\partial_\xi\A_z),\label{eq:Abphibar}\\
\bar b_z&=& \delta((3-a)A_\phi-Z\partial_Z\A_\phi+\partial_\xi\A_\phi-q\partial_\xi\A_r).\label{eq:Abzbar}
\eea

The equations in $\bar{\bf A}$ can be written much more simply in terms of these quantities. They become three first order partial differential equations in the variables $\xi$ and $Z$ in the form;
\bea
\bar b_r\Delta+\partial_Z\bar b_\phi-w\bar b_\phi+v\bar b_z-q\partial_\xi\bar b_z&=&0,\label{eq:AtwovariablefieldR}\\
w\bar b_r-\partial_Z\bar b_r+\bar b_\phi\Delta+\partial_\xi \bar b_z+\bar b_z(1-a-u)-Z\partial_Z\bar b_z&=&o,\label{eq:Atwovariablefieldphi}\\
q\partial_\xi\bar b_r-v\bar b_r-\partial_\xi\bar b_\phi+\bar b_\phi(u+a-2)+Z\partial_Z\bar b_\phi+\bar b_z\Delta&=&0.\label{eq:AtwovariablefieldZ}
\eea

 If one makes the same Fourier ans\"atz for the $\xi$ dependence as above ($\bar{\bf b}=\tilde{\bf b}\exp{ip\xi}$) we find 
 three first order ordinary equations to be solved for $\tilde{\bf b}(Z)$, which is a complex vector function of a real variable.   However a semi-analytic solution still requires the neglect of terms in $Z$.  After this approximation the fields are given in terms of the variable $\kappa$ of the text plus the appropriate coefficients required for a non-trivial solution.  Some special exact solutions may be found  using these equations. A disadvantage  is that it is not evident that the divergence  of the magnetic field remains zero, but these may be found; such as the example referred to in the text.

In this first attempt to explain certain observations in edge-on galaxies we continue with the vector potential approach to the magnetic fields , which guarantees a solenoidal magnetic field. We always work at small cone angle (measured from the equator) so that $Z$ is small. We then neglect all terms proportional to Z in the equations used in the text. We also restrict ourselves to the case where $a=2$, which `class' corresponds to an invariant quantity with the Dimensions of specific angular momentum. This allows us to seek a modal solution with combined $\xi$ and $Z$ dependences in the form 
\be
{\bf\ A}={\bf C}\exp(ip\kappa),\label{eq:Akappa}
\ee
where $p$ and $\kappa$ are as in the text. We must keep $Z$ small however.

There is one additional assumption that greatly simplifies the proceedings, but may not be strictly justified. In the third line of the first equation in this appendix we find the term $\partial_Z(Z\partial_Z\A_z)$. There is a term proportional to $\partial_Z\A_z$ after the first differentiation.  We neglect this term
compared to $\partial_\xi\partial_Z\A_z$ on the grounds that asymmetric magnetic fields should vary strongly with $\xi$.  Given our ans\"atz, this holds to the extent that $p>1$.   This assumption is consistent with neglecting terms proportional to $Z$ in the magnetic field equations.  Retaining this term complicates the calculations and changes them in detail, but the general behaviour remains similar.

 \bigskip\bigskip
 \appendix
 \centerline{\bf Appendix B :Kalnajs Gravitational Spiral Arms}
 In this appendix we calculate the type of scale invariant gravitational spiral arm that may be associated with the magnetic scale invariant arm. The gravitational field is shown in and above the plane. An initial overlay of corresponding modes shows that the magnetic arm lies along the gravitational arm. However the velocities in the pattern frame are different, which should lead to separation of the gravitational arm in the sense of gravitational arm leading at large radius. the reverse may be true at small radius.
 
The gravitational potential due to massive logarithmic spiral arm modes was found in \citet{Kal1971} and was discussed succinctly in \citet{BT2008}, page 107 et seq., and their problem (2.19). In \citet{Hen2011} the Poisson integral procedure used in these arguments was shown to be a scale invariant procedure, having a similarity class $a=5/4$. However the Poisson integral does not give the gravitational field above the plane. Such a field  is of possible use in discussing the influence of disc outflow on the dynamo, so we give a more standard procedure here that emphasizes the scale invariance.

 We begin with the Laplace equation in spherical coordinates $\{r,\theta,\phi\}$.  We seek scale invariant solutions by writing 
\be
\delta r=e^{\delta R}, ~~~~\Phi=\Psi(\xi,x)e^{2(1-a)\delta R},~~~~\sigma=\Sigma(\xi)e^{(1-2a)\delta R},\label{eq:BSS}
\ee
where $\Phi$ is the gravitational potential (with Dimension equal to velocity squared- see equation (\ref{eq:vformsum}) of the text), $x\equiv \cos{\theta}$ and $\xi$ is as defined in equation (\ref{eq:logspiral}) of the text. We have again defined $a\equiv \alpha/\delta$, but we should note that the scaling is along the spherical radius here, rather than along the cylindrical radius. In the expression for the surface density, we have eliminated the mass Dimension by using the invariant Dimensions of Newton's constant (e.g. \citet{Hen2015}).

Under these transformations the Laplace equation becomes 
\be
[1+\frac{q^2}{1-x^2}]~\partial_\xi^2\Psi+(5-4a)~\partial_\xi\Phi+\partial_x[(1-x^2)\partial_x\Psi]+2(3-2a)(1-a)\Psi=0.\label{eq:BscaleLaplace}
\ee
The equation is only separable for the similarity class $a=5/4$, whereupon the solution for the potential that is well-behaved in $x~~\epsilon~ [-1,1]$ is 
\be
\Psi=\sum_m~A_m~\exp{(i\frac{m}{q}\xi)}~P^m_\nu(x).\label{eq:BLsol}
\ee
Here $P$ is the associated Legendre function, $\nu\equiv -1/2\mp im/q$, and $A_m$ is a modal constant. We will normally use the $-$ sign in the indicated option.

The boundary condition at the galactic disc $(1/r)\partial_x\Phi|_o=2\pi G\sigma$ becomes 
\be
\partial_x\Psi|_o=\frac{2\pi G}{\delta}\Sigma(\xi),\label{eq:BBC}
\ee
from which we must determine $\{A_m\}$. From \citet{GR1994}, page 1026 we find
\be
\frac{dP^m_{~\nu}}{dx}|_o=\frac{2^{m+1}\sin{(\frac{(\nu+m)\pi}{2})}\Gamma(1+\frac{(\nu+m)}{2})}{\sqrt\pi\Gamma(\frac{(\nu-m+1)}{2})},
\ee

which can be combined with (\citet{AS1972}, p256)
\be
\sin{(\pi\frac{Z}{2})}=\frac{\pi}{\Gamma(\frac{Z}{2})\Gamma(1-\frac{Z}{2})}.\label{eq:Bident1}
\ee
Manipulating the result further using the identities
\be
\Gamma(1+Z)=Z\Gamma(Z),~~~~\Gamma(1-Z)=-Z\Gamma(-Z),
\ee
one obtains finally
\be
\frac{dP^m_{~\nu}}{dx}=-\frac{2^{(m+1)}\sqrt{\pi}}{\Gamma(-\frac{(\nu+m)}{2})\Gamma(\frac{1}{2}+\frac{(\nu+m)}{2})}.\label{eq:BPderiv}
\ee
This result is easily shown to agree with the result from MAPLE when their cut is taken from $1$ to $\infty$.

In principle we may now Fourier transform equation (B\ref{eq:BLsol}) to obtain $A_m$, however taking the density to be a pure spiral node $\Sigma(\xi)=\Sigma_m\exp{i(\frac{m}{q}\xi)}$ we obtain immediately that 
\be
A_m=-\frac{2\pi G\Sigma_m}{\delta}\frac{\Gamma(-\frac{(\nu+m)}{2})\Gamma(\frac{1}{2}+\frac{(\nu+m)}{2})}{2^{(m+1)}\sqrt{\pi}}.\label{eq:Bgravcoeff}
\ee
This completes the solution above the plane as given in equation (B\ref{eq:BLsol}). However to find the solution in the plane we need the value of $P^m_{~\nu}(0)$ which can be found in (\citet{GR1994}, p1026). 
The result for the modal logarithmic spiral arm in the plane is 
\be
\Psi_m(0)=-\frac{\pi G\Sigma_m}{\delta}\exp{(i\frac{m}{q}\xi)}\frac{\Gamma(-\frac{(\nu+m)}{2})\Gamma(\frac{(1+\nu-m)}{2})}{\Gamma(\frac{1-(\nu+m)}{2})\Gamma(1+\frac{(\nu-m)}{2})}.\label{eq:Bpotform}
\ee

Finally we  substitute  for $\nu$ and use the property $\Gamma(Z^*)=(\Gamma(Z))^*$, the asterisk indicating complex conjugate, to find
\be
\Psi(0)=-\frac{\pi G\Sigma_m}{\delta}\exp{(i\frac{m}{q}\xi)}\frac{|\Gamma(\frac{1}{4}-\frac{m}{2}\mp\frac{im}{2q})|^2}{|\Gamma(\frac{3}{4}-\frac{m}{2}\mp\frac{im}{2q})|^2}\equiv -\frac{G\Sigma_m}{\delta}\exp{(i\frac{m}{q}\xi)}N(m,q),\label{eq:Bpot}
\ee
which defines the modal coefficient $N(m,q)$. This coefficient is normally expressed (e.g. \citet{Kal1971}) in terms of $1-Z_1$ and $1-Z_2$ ($Z_1$ and $Z_2$ are the arguments in the definition of $N(m,q)$) which form can be shown to be identical to that of (B\ref{eq:Bpot}) by using the identity (B\ref{eq:Bident1}) plus the identity
\be
|\sin{Z}|^2=(\sin{x})^2+(\sinh{y})^2. 
\ee
Explicitly, the definition of $N(m,q)$ above becomes equal to  
\be
N(m,q)=\frac{|\Gamma(\frac{1}{4}+\frac{m}{2}\pm\frac{im}{2q})|^2}{|\Gamma(\frac{3}{4}+\frac{m}{2}\pm\frac{im}{2q})|^2}.\label{eq:BKalpotcoeff}
\ee
The two equivalent forms show the coefficient $N(m,q)$ to be invariant under a change in sign of $m$.
Normally the upper sign is used in each expression for $N(m,q)$. \footnote{This evaluates the integral that appears in the Poisson integral evaluation of the surface potential.}

This solution for the potential allows us to construct scale invariant, gravitational log spirals, that are fixed in a uniformly rotating (pattern) reference frame . Normally these spiral arms are superimposed on a smooth disc background (e.g. a Mestel disc) plus an isothermal dark halo (e.g. \citet{HI2016}). For the moment we consider this background to be in equilibrium and regard the spiral arms as a perturbation.

 We can thus expect the spiral arm gravitational potential to drive a magnetohydrodynamic flow in the rotating pattern frame, which can  interact with the log spiral dynamo arm. The gravitational acceleration produced by a given arm in and above the plane, varies as $1/(\delta r)^{3/2}$ near a given value of $\xi$ when $a=5/4$. Assuming the free-fall time to be approximately  $\sim\sqrt{r/G\Sigma_m}$ at $r$, one might therefore expect free-fall velocities to vary as $1/\delta r$. This dependence is what we require in the scale invariant steady dynamo of class $a=2$.  One has to ignore gas pressure, magnetic pressure and tension, and centrifugal/Coriolis forces in the rotating pattern frame on the free fall time scale. Such forces may also enforce a static equilibrium.  

%It may appear strange that the steady gravitational-arm similarity class ($5/4$) is different from that of the steady dynamo we study, namely $2$. However they are independent physical systems in the pattern frame, except through velocity coupling. The velocites implied by the gravitational arm (in free fall) do seem to be broadly compatible with the dynamo scale invariant form. 

%The scaling is isotropic for the gravitational arm in spherical coordinates, but it is  possibly anisotropic in the cylindrical dynamo scaling. We have used $a=\alpha/\delta$ as the definition of class in both cases without distinguishing $\delta$. Suppose however that we choose $a=\alpha/\delta_\perp $ for the cylindrical dynamo class while leaving the gravitational class the same. By taking $\alpha/\delta=2$ we require $\alpha/\delta_\perp=1+\delta_\parallel/\delta_\perp$ (remembering $\delta=(\delta_\perp+\delta_\parallel)/2$). Hence we would obtain the same class as for the gravitational potential if $\delta_\parallel/\delta_\perp=1/4$.  This implies that the vertical scale height would be roughly one quarter of the disc radial scale. However the principal point of this  argument is that we should not worry too much about discordant gravitational and dynamo classes, particularly when different coordinates are used.

\begin{figure}%[p]
\begin{tabular}{cc} %This will make a one-column figure
\rotatebox{0}{\scalebox{0.4} %change the angle and scale as you need
{\includegraphics{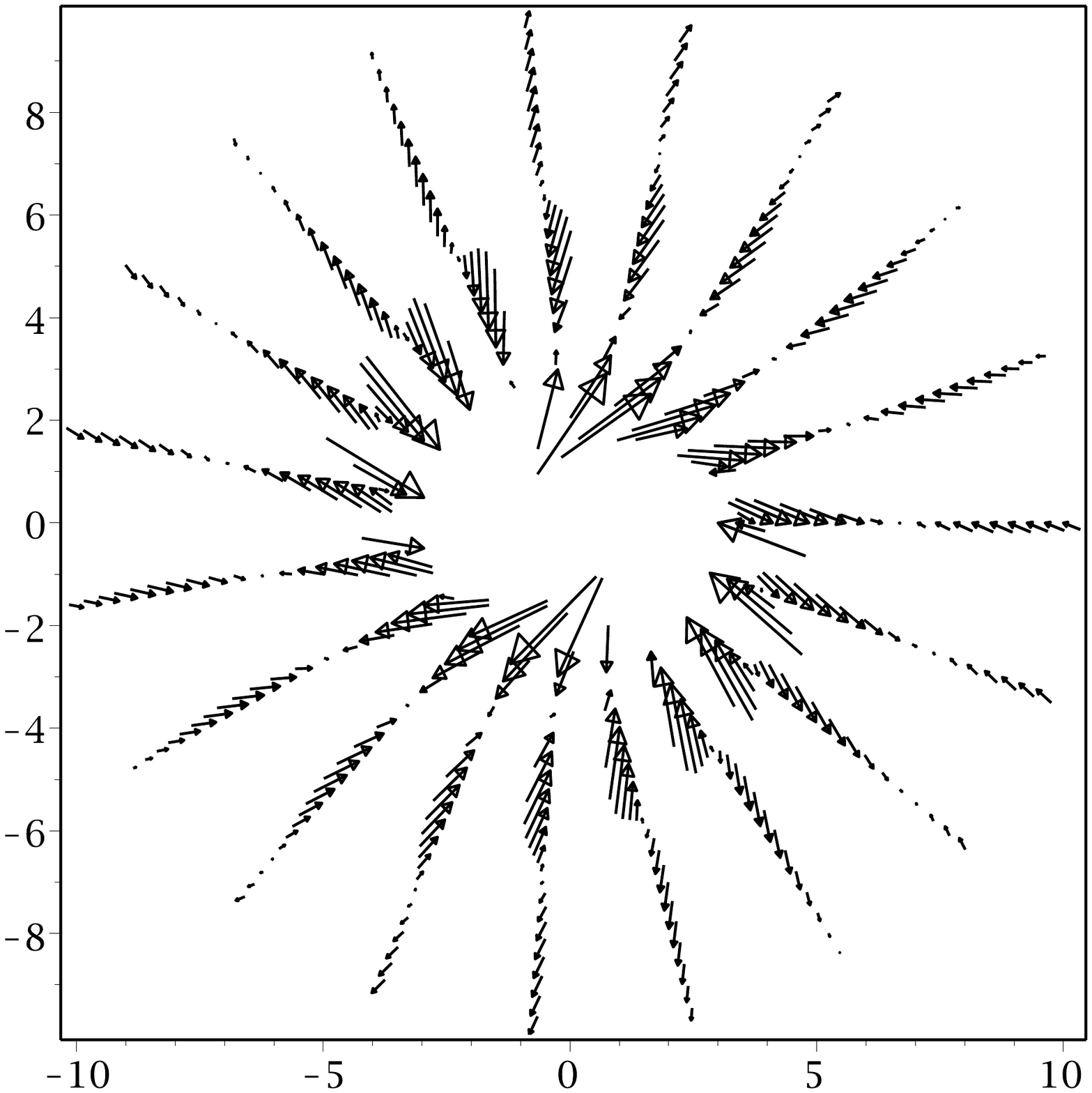}}}& 
\rotatebox{0}{\scalebox{0.4} %change the angle and scale as you need
  {\includegraphics{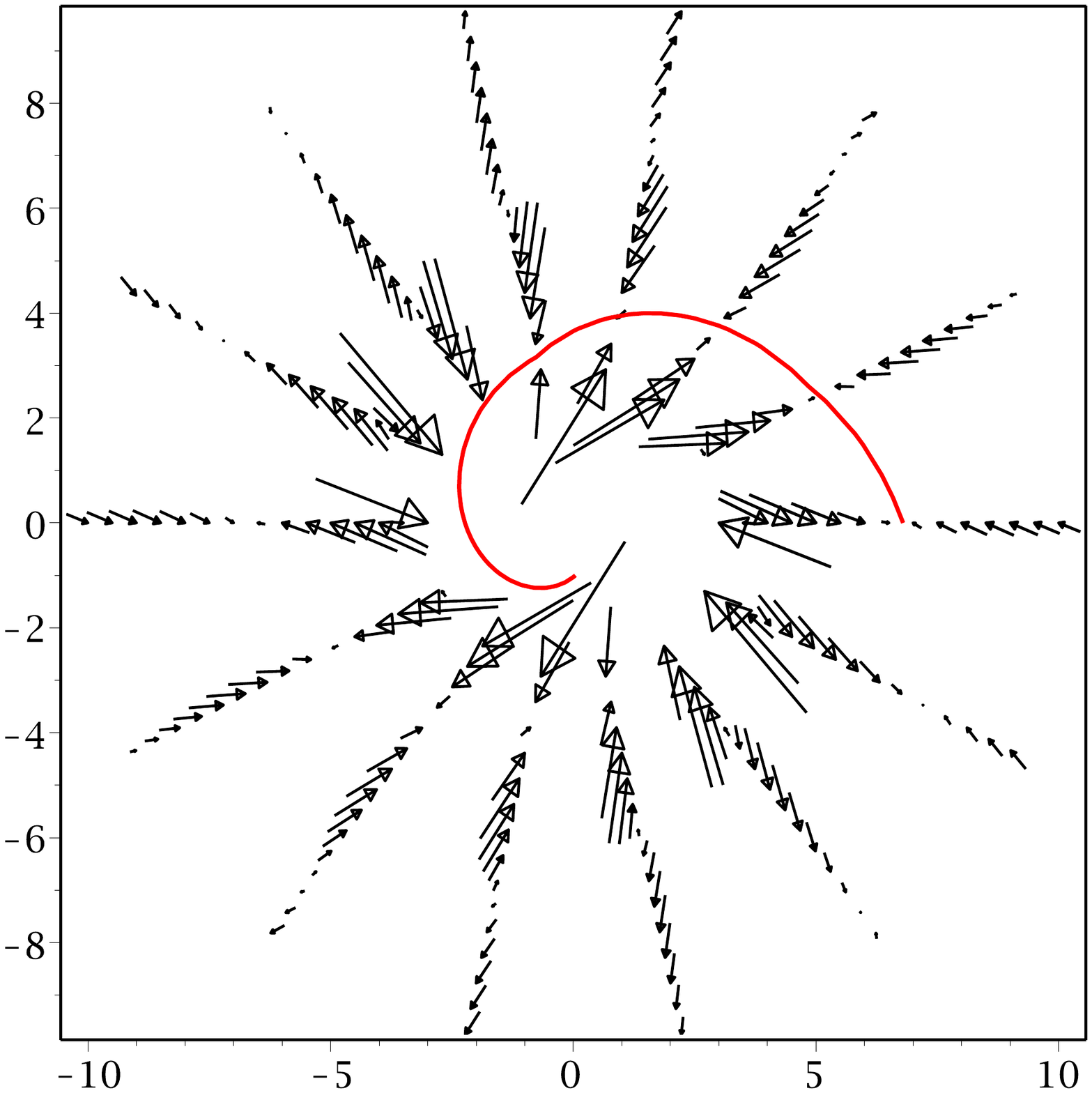}}}\vspace{-8.5mm} \\  \vspace{-7.0mm}
{\rotatebox{0}{\scalebox{0.4} %change the angle and scale as you need
{\includegraphics{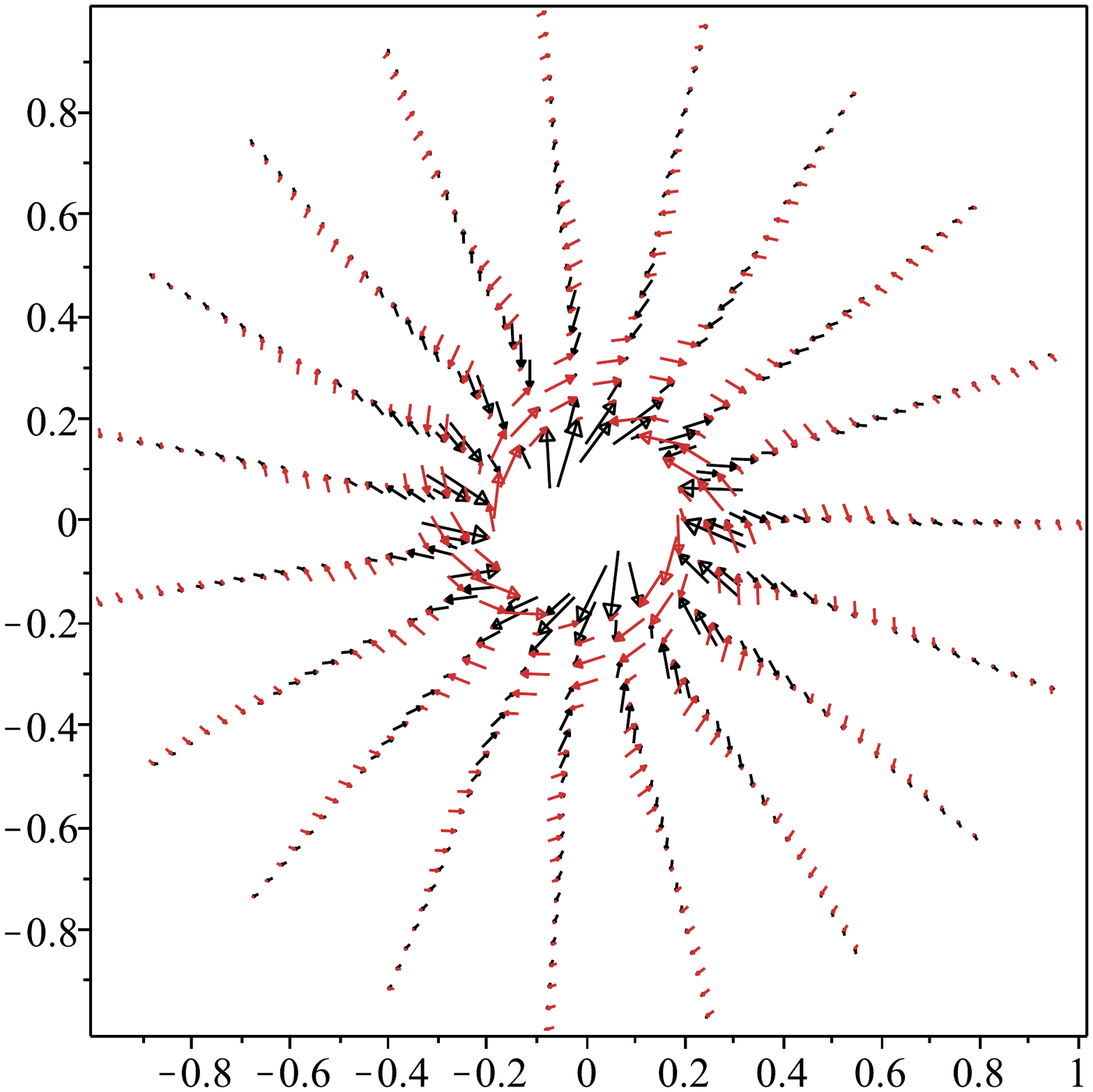}}}}&
\rotatebox{0}{\scalebox{0.9} %change the angle and scale as you need
{\includegraphics{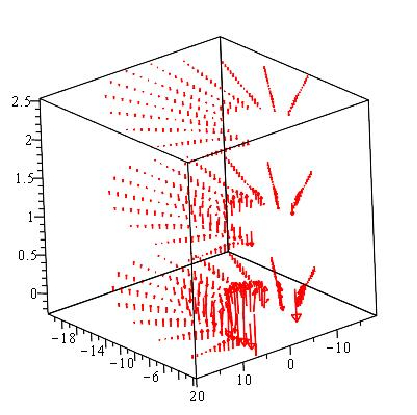}}}
\end{tabular}
\caption{The upper left figure shows the gravitational field in the plane of a single mode with $m=2$ and the pitch angle tangent $q=0.4$ (pitch angle $\sim 22^\circ$). The radius range is from $3$ to $10$ in Units of $1/\delta$ in all images except lower left. The gravitational field is in Units of $\delta/\pi G\Sigma_m$. Arrows are the local gravitational field at the arrow tip divided by the average, increased by $50$\% for clarity. The upper right figure shows a rough trace of the arm peak at $\xi\approx 6.7$.
The vectors are the local gravitational field divided by the average and reduced by $50$\%.  At lower left the gravitational $m=2$ mode is combined  with the $m=-2$ magnetic mode. The image at lower right shows the force field directions of the arm defined at upper left. The radius runs from $6.7$ to $20$, while the azimuth runs from $-2.734$ to $0$ in order to trace out the spiral arm. }    
\label{fig:Bgravspiralfields}
\end{figure}
We show in figure (B\ref{fig:Bgravspiralfields}) the gravitational field of an $m=2$ mode with a pitch angle of $\sim 22^\circ$in the upper left image. The gravitational arm lies between the oppositely pointing heads of the gravitational field vectors. At upper right the continuous line traces the expected mass concentration of the spiral arm for emphasis. The figure at lower right shows the gravitational field vectors above the plane for the same parameters. The spiral arm can be traced by the distribution of downward pointing vectors. 

At lower left the same gravitational mode is combined with a $m=-2$ (actually $m=+2$ but $z=0^-$) magnetic mode. At this  stage the magnetic arms lis along the gravitational arms. The  azimuthal speed of the gravitational arm relative to the pattern frame should vary as $g(\xi)r^{-(1/4)}$, while that of the magnetic arm should vary as $f(\xi)/r$ due to the respective  scale invariance classes. Here $f$ and $g$ are unknown unless the azimuthal velocity of a given  arm  relative to the pattern is measured at a given  cylindrical radius.  Nevertheless one can speculate that a gravitational arm may move ahead of the magnetic arm at large radius so that the magnetic arm lies inside the gravitational arm. The reverse should be true at a sufficiently small radius. Hence the  magnetic and gravitational arms may cross at some radius. This is only likely if the velocities relative to the pattern are reasonably close to one another at some disc radius $(f/g)^{4/3}$ .

  %The lower left figure illustrates a well-defined delta function (in $\xi$) gravitational arm and corresponding field. One hundred modes with the same amplitude $\Sigma_m$ and smaller pitch angle ($\sim 14^\circ$) have been summed. There is a quite noticeable oscillation in the direction and strength of the gravitational field along the arm, which may suggest vortical instability associated with the formation of molecular clouds. In each field image the mass arm should lie between the oppositely directed arrows that indicate the gravitational field on a constant value of $\xi$. This  is indicated at upper right for $\xi=6.7$.

%\begin{figure}%[p]
%\begin{tabular}{cc} %This will make a two-column figure
%\rotatebox{0}{\scalebox{0.8} %change the angle and scale as you need
%{\includegraphics{spacecurveWm2const.eps}}}
%\rotatebox{0}{\scalebox{0.3} %change the angle and scale as you need
%{\includegraphics{.eps}}}\\
%{\rotatebox{0}{\scalebox{0.8} %change the angle and scale as you need
%{\includegraphics{spacecurveWm-2cons.eps}}}}
%\end{tabular}
%\caption{ The left-hand image shows the  magnetic field line originating at $m=2$, $r=2$ and $\phi=\pi$ in the disc. The right-hand image shows the magnetic field line originating in the disc from the same radius, with azimuth $\phi=\pi/12$,  but with $m=-2$. The scales are constrained to be equal on all three axes.   }    
%\label{fig:spacecurvesconstrained}
%\end{figure

\label{lastpage}
\end{document}